\documentclass[sigconf,natbib=true,anonymous=False]{acmart}
\usepackage{bm}
\usepackage{multirow}
\usepackage{subfigure}
\usepackage{algorithmic}
\usepackage[ruled, linesnumbered]{algorithm2e}
\usepackage{booktabs} 
\usepackage{amsmath}
\usepackage[normalem]{ulem}
\useunder{\uline}{\ul}{}

\usepackage{amssymb}
\usepackage{xcolor}
\usepackage{multirow}
\usepackage{graphicx}
\usepackage{subfigure}
\usepackage{amsfonts}
\usepackage{balance}
\usepackage{enumitem}
\setlist[itemize]{leftmargin=*}
\usepackage{xcolor, colortbl}
\usepackage{hyperref}
\usepackage{pifont}

\newtheorem{thm}{Theorem}

\newtheorem{remark}{Remark}
\newtheorem{cor}{Corollary}

\newcommand{\shortname}{\textsf{SEA}}
\newcommand{\oursfull}[0]{\textbf{\underline{S}}eparate L\textbf{\underline{e}}\textbf{\underline{a}}rning\xspace}

\hypersetup{
    colorlinks=true,            
    linkcolor=red,              
    citecolor=green,            
    filecolor=green,            
    urlcolor=magenta            
}
\begin{document}

\title{It is Never Too Late to Mend: Separate Learning for Multimedia Recommendation}

\author{Zhuangzhuang He}
\affiliation{%
  \institution{Hefei University of Technology}
  \city{Hefei}
  \country{China}
}
\email{hyicheng223@gmail.com}

\author{Zihan Wang}
\affiliation{%
  \institution{Hefei University of Technology}
  \city{Hefei}
  \country{China}
}
\email{zhwang.hfut@gmail.com}

\author{Yonghui Yang}
\affiliation{%
  \institution{Hefei University of Technology}
  \city{Hefei}
  \country{China}
}
\email{yyh.hfut@gmail.com}

\author{Haoyue Bai}
\affiliation{%
  \institution{
  Arizona State University}
  \city{Tempe}
  \country{USA}
}
\email{baihaoyue621@gmail.com}

\author{Le Wu}
\affiliation{%
  \institution{Hefei University of Technology}
  \department{Institute of Dataspace,}
  \institution{Hefei Comprehensive National Science Center}
  \city{Hefei}
  \country{China}
}
\email{lewu.ustc@gmail.com}

\begin{abstract}
Multimedia recommendation, which incorporates various modalities~(e.g., images, texts, etc.) into user or item representation to improve recommendation quality, and self-supervised learning carries multimedia recommendation to a plateau of performance, because of its superior performance in aligning different modalities. However, more and more research finds that aligning all modal representations is suboptimal because it damages the unique attributes of each modal. These studies use subtraction and orthogonal constraints in geometric space to learn unique parts. However, our rigorous analysis reveals the flaws in this approach, such as that subtraction does not necessarily yield the desired modal-unique and that orthogonal constraints are ineffective in user and item high-dimensional representation spaces.

To make up for the previous weaknesses, we propose~\oursfull~(\shortname) for multimedia recommendation, which mainly includes mutual information view of modal-unique and -generic learning. Specifically,  we first use GNN to learn the representations of users and items in different modalities and split each modal representation into generic and unique parts. 
We employ contrastive log-ratio upper bound to minimize the mutual information between the general and unique parts within the same modality, to distance their representations, thus learning modal-unique features.
Then, we design Solosimloss to maximize the lower bound of mutual information, to align the general parts of different modalities, thus learning more high-quality modal-generic features. 
Finally, extensive experiments on three datasets demonstrate the effectiveness and generalization of our proposed framework. The code is available at~\href{https://github.com/bruno686/SEA}{\underline{\shortname}} and the~\href{https://github.com/bruno686/SEA/tree/main/main_experiment_log}{\underline{full training record}} of the main experiment.

\end{abstract}


\maketitle

\section{Introduction}

%

\begin{figure*}[h!]
\centering
\includegraphics[width=0.9\textwidth]{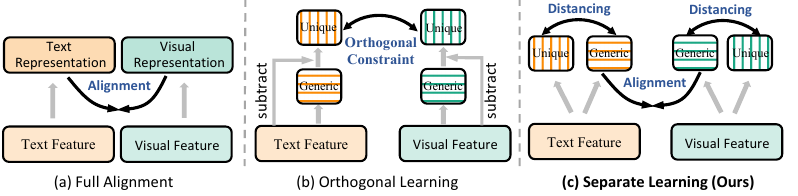} 
   \caption{We illustrate the difference between the three strategies, (a) The full alignment modality strategy of most SSL-based methods, (b) Learning modal-unique using orthogonal constraint. (c) Our separate learning strategy.}
\vspace{-5mm}
\label{intro}
\end{figure*}

With the development of multimedia platforms (e.g., Instagram, Twitter, etc.), users easily express their emotions with various modalities (images, text, etc.). Therefore, multimedia recommendation~\cite{he, bai, cai1, cai2, wang1, wang2, hui1, peijie1, peijie2} has emerged, which can deeply understand user preferences by exploiting multiple modalities. Early methods like VBPR~\cite{vbpr} added item modal features for better performance, and later, GNN-based methods~\cite{dualgnn, freedom, LATTICE, MMGCN} goes further. Building on these successes, self-supervised learning~(SSL)~\citeN{ssl-nlp-survey,ssl-rec-survey} carries multimedia recommendation to a new plateau of performance. The SSL-based methods have the advantage of both exploring potential relationships between modalities and preventing preference shortcuts~\cite{MMSSL} from unimodal, yielding more expressive representation. Specifically, most studies~\cite{BM3, MMSSL, mentor} use SSL (e.g. contrastive learning~\cite{cl}) to full alignment representation of each modality~(Figure~\ref{intro} (a)).

However, more and more study~\cite{spe-common1,spe-common2, PAMD,spc-common4, MGCN} note that aligning modality is good, but it should not be the entirety. 
Because textual and visual modalities both describe an item's color and shape, images provide unique depth and texture, while text offers syntactic and morphological details~\cite{SimMMDG}. 
Therefore, most recent work~\cite{spe-common1,spe-common2,PAMD,spc-common4, MGCN} mainly focuses on learning modal-unique features, which is important to improve the quality of multimedia representation and recommendation performance~\cite{MICRO,MGCN,SimMMDG}.
A popular obtaining modal-unique representation paradigm is orthogonal learning ~\cite{spe-common1,spe-common2,PAMD,spc-common4,MICRO,MGCN}, which first usually obtains a modal-generic by encoders, and then subtract the generic part from the entire to obtain a modal-unique. 
Finally, they apply orthogonal constraints to the unique parts of different modalities to learn the unique attributes of each modality as shown in Figure~\ref{intro}~(b).

Despite their notable performance, we argue that orthogonal learning paradigm is ineffective. First of all, we clarify what the definition of modal-unique is. According to the most previous definition~\cite{spe-common1,spe-common2, PAMD,spc-common4, MICRO, MGCN}, it refers to attributes that are unique within each modality and that are distinct from the modal-generic.
So as shown in Figure 1 (b), can we obtain modal-unique by subtraction? 
Even if this holds, then secondly, can we ensure that unique attributes of each modality are learned by imposing orthogonal constraints on the unique parts of the different modalities?
For the first point, we consider if the modal-generic is not adequately learned, then the modal-unique will be impeded, leading to cumulative errors. For the second point, we get that pairs of vectors tend to be orthogonal by deriving the angle of any vector in a high-dimensional space, so the constraint is invalid. Moreover, since the goal is to learn attributes distinct from the modal-generic parts, it is unreasonable to impose learning constraints on the unique parts of different modalities.
Specific detailed analyses of these two questions are in~\hyperref[analysis]{Motivation}.

To make up for the previous weaknesses, we propose~\oursfull~(\shortname) for multimedia recommendation, which mainly includes a mutual information view of modal-unique and -generic learning.
First, we use GNN to obtain multimodal representations. Then, we split each modal feature into two parts to learn modal-generic and model-unique. We argue that within a modality, the unique and general parts should be as different as possible, while the general parts across different modalities should be as consistent as possible. 
Specifically, we employ contrastive log-ratio upper bound to minimize the mutual information between the general and unique parts within the same modality, to distance their representations, thus learning modal-unique features.
Then, we design Solosimloss to maximize the lower bound of mutual information, to align the general parts of different modalities, thus learning more high-quality modal-generic features. 
To summarize, our contributions are as follows:
\begin{itemize}
  \item 
  We pioneer efforts in an insightful analysis of the flaws in the orthogonal learning paradigm for obtaining modal-unique information in multimedia recommendation.
  \item 
  We design a separate learning framework~\shortname. We directly split modal-unique and modal-generic to avoid accumulated errors. And, one of our cores is that we minimize the upper bound of mutual information between the unique and generic parts within the same modality to learn modal-unique representation.
  \item
  After empirical experiments on several datasets and various analytical validations,~\shortname~is well demonstrated in terms of effectiveness, generalization, and flexibility.
\end{itemize}

\section{Motivation: Is the Current Paradigm Ideal?}
\label{analysis}
Obtaining modal-unique representations using a popular orthogonal learning paradigm~\cite{spe-common1,spe-common2,PAMD, MGCN, MMSSL} involves the two steps shown in Figure~\ref{intro} (b).

\noindent (1) \textit{Subtract the modal-generic feature to get modal-unique feature.} They first obtain the modal-generic $\boldsymbol{h}_{g}$~\footnote{Notation: M represents modalities, including multiple m, such as t (text), and v (visual). $\boldsymbol{h}$ is representation, g is generic in subscripts and u is unique.}.
Then, they subtract from the full representation of each modality to obtain modal-unique representation
\begin{small}              
$\boldsymbol{h}^{\mathrm{v}}-\boldsymbol{h}_{g}$ $(\boldsymbol{h}^{\mathrm{t}}-\boldsymbol{h}_{g}).$
\end{small}

\noindent (2) \textit{Imposing orthogonal constraints on unique parts of different modalities ensures that the unique features of each modality are learned.} They apply orthogonal constraints
\begin{small}
$\Vert\boldsymbol{h}_{u}^{\mathrm{v}}{ }^{\top} \boldsymbol{h}_{u}^{\mathrm{t}}\Vert^2$.
\end{small}
to the unique parts 
\begin{small}    
$\boldsymbol{h}^\mathrm{v}_{u} (\boldsymbol{h}^\mathrm{t}_{u})$ 
\end{small}
of different modalities to ensure that the learned representations are complementary to the modal-generic.

Now let us get back to the original intent. We aim to obtain information that is unique within each modality and that is distinct from the modal-generic. The core of the previous works lies in subtraction and learning modal-unique between different modalities using orthogonal constraints. For this, we raise two questions.

\noindent \ding{172}~\textit{Can we obtain modal-unique by subtraction?} In high-dimensional space, data relationships are often complex and nonlinear. Linear models (e.g., obtaining unique information through subtraction) assume linear relationships between features, which is often not the case. Data may lie on complex manifolds, and linear assumptions fail to capture these relationships effectively. In addition, if the modal-generic is not adequately learned, then the modal-unique will be impeded, leading to cumulative errors.

\noindent \ding{173}~\textit{Can we ensure that unique attributes of each modality are learned by imposing orthogonal constraints on the unique parts of the different modalities?} Firstly, in mathematics, two vectors being orthogonal means their inner product is zero. Orthogonality emphasizes geometric independence, not informational independence. And the modal-unique should encompass meaningful content. Moreover, vectors are often orthogonal in high-dimensional space, even without orthogonality constraints. In addition, we give a theoretical analysis in theorem~\ref{them1}. Secondly, considering that our goal is to learn unique attributes from each modality, which are distinct from the modal-generic attributes, we argue it is not reasonable to use the unique parts of different modalities as the constraint target.
{\thm{Suppose two random vectors $x$ and $y$ in n-dimensional space, which are at an angle $\theta$. $x$ and $y$ are almost orthogonal in general high-dimensional space. \label{them1}}}
\begin{equation}
p_n(\theta)=\frac{\Gamma\left(\frac{n}{2}\right)}{\Gamma\left(\frac{n-1}{2}\right) \sqrt{\pi}} \sin ^{n-2} \theta.
\end{equation}
{\remark{From $p_n(\theta) \sim \sin ^{n-2} \theta$, it can be found that the maximum probability is $\theta=\frac{\pi}{2}$ when $n \geq 3$, and also that $\sin ^{n-2} \theta$ is symmetric about $\theta=\frac{\pi}{2}$, so its mean is also $\frac{\pi}{2}$. In multimodal recommendation, the user, item dimension is usually set above 32~\cite{MMGCN, MMSSL,mmrec}. The full proof is in~\ref{proof1} (Eq.20 - Eq.25). In addition, we want to make sure that the angle is stabilized at x for n greater than 3, so we give the following definition of the variance of n.}}

{\cor{The Theorem 1 describes the distribution adequately, and we need to consider the variance of $\theta$ to ensure that the angle is stabilized at $\frac{\pi}{2}$ when n is relatively large.}}
\begin{equation}
\operatorname{Var}_n(\theta)=\frac{\Gamma\left(\frac{n}{2}\right)}{\Gamma\left(\frac{n-1}{2}\right) \sqrt{\pi}} \int_0^\pi\left(\theta-\frac{\pi}{2}\right)^2 \sin ^{n-2} \theta d \theta.
\end{equation}
From this formula, we can derive the following approximate analytical solution for $\theta$,
\begin{equation}
\sin^{n-2} \theta \approx \exp \left[-\frac{n-2}{2}\left(\theta-\frac{\pi}{2}\right)^2\right].
\end{equation}

{\remark{From this approximation, we can roughly consider that \(\theta\) follows a normal distribution with mean \(\frac{\pi}{2}\) and variance \(\frac{1}{n-2}\). In other words, as \(n\) becomes larger, the variance approximately becomes \(\frac{1}{n-2}\), indicating that the variance decreases as \(n\) increases. The full proof is in~\ref{proof1} (Eq.26 - Eq.28)}. This rigorously demonstrates that vectors are naturally orthogonal in high-dimensional spaces.}

\noindent \textbf{Summary.} We analyze the flaws of existing paradigm for obtaining modal-unique: \textit{the dependency of modal-unique on modal-generic}, \textit{the inadequacy of constraint objects}, and \textit{the ineffectiveness of orthogonal constraints}. 


\section{Task Description}
We give a formal definition of the multimedia recommendation, which focuses on using multiple modalities $m$, including text $t$, visual $v$. First, we give some notational conventions~\cite{mmrec,MMSSL,MMGCN}.
We use $u \in \mathcal{U}$ to represent user and $i \in \mathcal{I}$ to represent item. For user and item embedding, $\mathbf{E}_u^f \in \mathbb{R}^{d_f \times |\mathcal{U}|} $ represents the user's randomly initialized free embedding, $\mathbf{E}_i^v \in \mathbb{R}^{d_v \times |\mathcal{I}|}$ represents item initialized embedding with image features, and $\mathbf{E}_i^t \in \mathbb{R}^{d_t \times |\mathcal{I}|}$ represents item initialized embedding with text features.
$A$ is the adjacency matrix of a user-item bipartite graph,  In addition, $\mathcal{R} \in \{0, 1\}^{|\mathcal{U}| \times|\mathcal{I}|}$ denotes user interaction history, in which entry $\mathcal{R}_{u, i}=1$ represents a user interacting with an item, while $\mathcal{R}_{u, i}=0$ represents a user not interacting with an item. We predict the rating $\hat{r}_{u, i}$ by graph encoders $G$. The optimization objective for multimedia recommendation is as follows:
\begin{equation}
\label{problem1}
   \theta^*=\underset{\theta}{\arg \min }~\mathbb{E}_{(u, i) \sim \mathcal{P}} \mathcal{L}_{\text {rec}}(G(A, \mathbf{E}_u^f, \mathbf{E}_i^v, \mathbf{E}_i^t),~\mathcal{R}), 
\end{equation}
where $\mathcal{P}$ denotes distribution of training data, $\mathcal{L}_{\text {rec}}$ is BPR~(Bayesian Personalized Ranking) loss function. $\theta$ is the total of the parameters of the model and the embedding parameters, and $\theta^*$ is optimal parameters. Furthermore, we build homogeneous item-item multimodal graphs to establish semantic correlations within modalities. Specifically, the modal features are clustered by $k$NN~($k$-Nearest Neighbor)~\cite{knn} to obtain the item-item visual connection matrix $S_v$ and textual $S_t$. Finally, we adjust Eq.~\eqref{problem1} as follows:
\begin{equation}
\label{problem2}
   \theta^*=\underset{\theta}{\arg \min }~\mathbb{E}_{(u, i) \sim \mathcal{P}} \mathcal{L}_{\text {rec}}(G(A, \mathbf{S}_v, \mathbf{S}_t, \mathbf{E}_u^f, \mathbf{E}_i^v, \mathbf{E}_i^t),~\mathcal{R}).
\end{equation}

\section{The Proposed \shortname~Framework}
\begin{figure*}
\centering
    \includegraphics[width=0.8\textwidth]{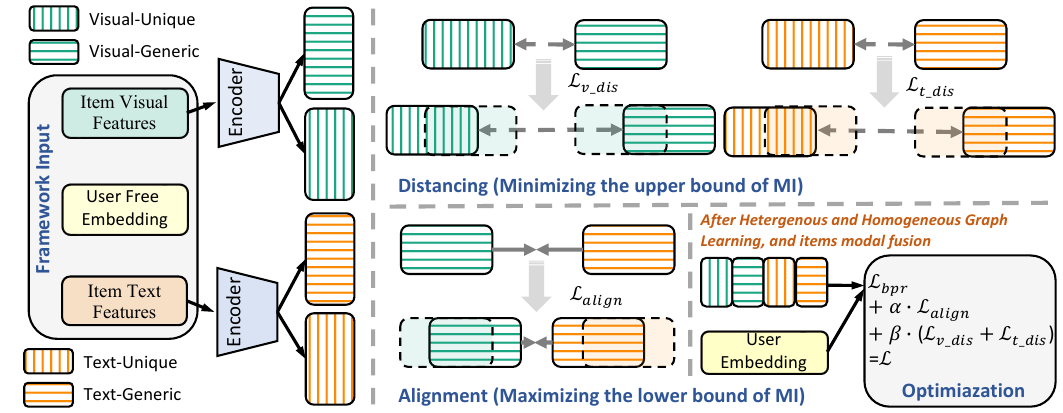} 
    \caption{Overall our proposed framework.} 
    \vspace{-2mm}  
    \label{Overall}
\end{figure*}

In~\hyperref[analysis]{Motivation}, we analyze the weaknesses of existing orthogonal learning paradigms for acquiring modal-unique in geometric spaces. Therefore, we propose separate learning for multimedia recommendation.
Specifically, two modules are designed, 1) GNN-based multimodal representation, and 2) mutual information view of modal-unique and -generic learning. 
The core idea is that we first divide the whole modal representation into modal-generic and modal-unique.
Then, modal-unique is obtained by minimizing the upper bound of the mutual information, and modal-generic is obtained by maximizing the lower bound of the mutual information. This is simple but more rigorous and theoretically guaranteed from an information-theoretic perspective.
The overall model framework is shown in Figure~\ref{Overall}.

\subsection{GNN-based Multimodal Representation}
In previous works~\cite{MICRO, freedom}, GNN-based multimedia recommendation has shown remarkable capabilities. We extend that advantage in our proposed framework by modeling two kinds of multimodal graphs using graph convolution networks, heterogeneous user-item graph, and homogeneous item-item graphs. These two kinds of graphs model the collaborative signals and the semantic associations between items.

\subsubsection{Heterogeneous Multimodal User-Item Graph.}
First, we construct two heterogeneous multimodal user-item graphs.
Then, we perform the graph convolution operation~\cite{lightgcn} to get user and item representation. Specifically, the convolution form is shown below:
\begin{equation}
\begin{aligned}
& {\mathbf{E}_u^m}^{(k+1)}=\sum_{i \in \mathcal{N}_u} \frac{1}{\sqrt{\left|\mathcal{N}_u\right|} \sqrt{\left|\mathcal{N}_i\right|}} {\mathbf{E}_i^m}^{(k)}, \\
& {\mathbf{E}_i^m}^{(k+1)}=\sum_{u \in \mathcal{N}_i} \frac{1}{\sqrt{\left|\mathcal{N}_i\right|} \sqrt{\left|\mathcal{N}_u\right|}} {\mathbf{E}_u^m}^{(k)},
\end{aligned}
\end{equation}
where $k$ denotes the convolution $k$ times, $m$ represents modalities, and $\mathcal{N}_u$ where comes from the row corresponding to $u$ in adjacency matrix $A$, which represents the one-hop neighbourhood of $u$ and vice versa for $i$.
We stack the representation of the $K$-layer convolutional network to get the final representation:
\begin{equation}
\mathbf{E}_u^m=\sum_{k=1}^K {\mathbf{E}_u^m}^{(k)} ; \quad \mathbf{E}_i^m=\sum_{k=1}^K {\mathbf{E}_i^m}^{(k)}.
\end{equation}

\subsubsection{Homogeneous Multimodal Item-Item Graph.}
To further enhance the expression of item representation, we follow~\cite{freedom, LATTICE} to model the semantic associations between items. First, we use $k$NN to model the connection $\mathbf{S}$ between items. Additionally, the initial representation of the graph is a fusion of the multiple modal representation~$\textbf{E}_i=\{\textbf{E}_i^v||\textbf{E}_i^t\}$ obtained above. Specifically, we calculate the two-by-two (i and j) cosine similarity between items, for example between item $i$ and $i^{\prime}$, as follows:
\begin{equation}
\mathbf{S}_{i, i^{\prime}}^m=\frac{\left(\mathbf{E}_i^m\right)^{\top} \mathbf{E}_{i^{\prime}}^m}{\left\|\mathbf{E}_i^m\right\|\left\|\mathbf{E}_{i^{\prime}}^m\right\|},
\end{equation}
where~$m$ stands for textual modality or visual modality. We then use $k$NN sparsification~\cite{freedom} to preserve top-$k$ similarity connections as edges of our item-item graph:
\begin{equation}
    \mathbf{S}_{i, i^{\prime}}^m=\left\{\begin{array}{ll}
    1, & \mathbf{S}_{i, i^{\prime}}^m \in \text {top-} k\left(\mathbf{S}_{i, i^{\prime}}^m\right), \\
    0, & \text { otherwise. }
    \end{array} \right.
\end{equation}
In addition, to prevent some vertices with high degrees and those with low degrees from having a large difference in the feature distribution, we perform a symmetric normalization for $\mathbf{S}^m$,
\begin{equation}
\mathbf{S}^m=\left(\mathbf{D}^m\right)^{-1/2} \mathbf{S}^m\left(\mathbf{D}^m\right)^{-1/2},
\end{equation}
where~$\mathbf{D}^m$ represents the diagonal degree matrix of $\mathbf{S}^m$.
We weight and sum the connection matrix of each modality~$v$ and $t$ to obtain the final matrix~$\mathbf{S}$ as follows:
\begin{equation}
    \mathbf{S} = \textbf{W}_s \cdot \mathbf{S}^v + (1-\textbf{W}_s) \cdot \mathbf{S}^t,
\end{equation}
where~$\textbf{W}_s$ is learnable weight matrix. We define the initial graph convolutional representation as $\textbf{H}^{(0)}=\textbf{E}_i$.
We then aggregate the neighborhood representation of item~$i$ to extract semantic associations as follows:
\begin{equation}
\mathbf{H}^{(k)}=\mathbf{S} \cdot \mathbf{H}^{(k-1)}.
\end{equation}
In summary, we obtain the final items representation~$\mathbf{H}$ of the homogeneous graph after $k$ times convolution.

\subsection{Mutual Information Perspective of Modal-Unique and -Generic Learning}
We learn modal-unique and modal-generic representations from the perspective of mutual information (MI). Unlike previous approaches~\cite{spe-common1,spe-common2,PAMD,spc-common4, MGCN} that use subtraction and orthogonal constraints, MI provides a more rigorous guarantee from the data distribution perspective. We first split each modal representation into two parts to learn modal-unique and -generic. 
Then, we argue that obtaining \textit{modal-unique} representations is equivalent to \textbf{distancing} the distribution of unique parts within the same modality.
Specifically, \ding{182} we~\uline{minimize the upper bound} of MI of the unique part of the same modality. 
In addition, we believe that obtaining \textit{modal-generic} representations is equivalent to \textbf{aligning} the distribution of generic parts of different modalities. Thus, \ding{183} we~\uline{maximize the lower bound} of MI of the generic part of the different modalities. For this, we design a Solosimloss, which has the advantage of enabling contrastive learning without the need for negative sampling, making it more efficient.

\begin{equation}
    \mathbb{I}(x ; y) \geq \mathbb{I}_{\mathrm{lower}}(x ; y); \qquad
    \mathbb{I}(x ; y) \leq \mathbb{I}_{\mathrm{upper}}(x ; y),
\end{equation}
where we give two forms of MI optimization.

\subsubsection{Splitting modal Representation}
Inspired by~\cite{SimMMDG}, we split items representation $\mathbf{E}_i^m$ into two parts and name them modal-unique~$\mathbf{E}_q^m$ and modal-generic~$\mathbf{E}_g^m$. After that, we describe how to learn both.

\subsubsection{Minimizing the Upper Bound for Modal-unique Learning}
Modal-unique representation reveals information that is distinctive to each modality, which is essential for enhancing the expression of modal representation~\cite{spe-common1,spe-common2,PAMD,spc-common4}. 
The previous approach is to perform subtraction and orthogonal constraints in geometric space to obtain modal-unique. We determined after a deep analysis that in geometric space is a suboptimal strategy.
From mutual information perspective, we expect the MI between generic and unique parts to be as small as possible to acquire modal-unique representation. Directly minimizing ~$\mathbb{I}(\mathbf{E}_g^m;~\mathbf{E}_q^m)$ is intractable, thus we convert MI minimization into exploring the upper bound, specifically, deriving a sample-based MI upper bound based on the Contrastive Log-ratio Upper Bound (CLUB)~\cite{club}.
For notational simplicity in subsequent reasoning, we rewrite that expression here as~$\mathbb{I}\left(\mathbf{E}^g;~\mathbf{E}^q\right)$.

Given that $\mathbf{E}_j^g \sim p\left(\mathbf{E}^g\right)$, if the conditional distribution $p(\mathbf{E}_i^g | \mathbf{E}_i^q)$ between $\mathbf{E}_i^g$ and $\mathbf{E}_i^q$ is known, then
\begin{equation}
  \mathbb{I}_{\mathrm{upper}}\left(\mathbf{E}_i^g;~\mathbf{E}_i^q\right) \mathrel{:=} \mathbb{E}\left[\log p\left(\mathbf{E}_i^g \mid \mathbf{E}_i^q\right)-\frac{1}{|\mathcal{I}|} \sum_{j=1}^{|\mathcal{I}|} \log p\left(\mathbf{E}_j^g \mid \mathbf{E}_i^q\right)\right]. 
\end{equation}

However, the computation of $p(\mathbf{E}_i^g \mid \mathbf{E}_i^q)$ is intractable. Thus, we approximate $p(\mathbf{E}_i^g \mid \mathbf{E}_i^q)$ with the variational distribution $q_\phi(\mathbf{E}_i^g \mid \mathbf{E}_i^q )$~\cite{club}. Subsequently, we employ log-likelihood maximization to iteratively learn the parameters $\phi$ accordingly. Consequently, the upper bound of MI can be expressed as follows:
\begin{equation}
\begin{aligned}
\label{dis_mi}
\mathcal{L}_{m\_\text{dis}} 
&= \frac{1}{|\mathcal{I}|} \sum_{i=1}^{|\mathcal{I}|} 
 \left[ \log q_\phi\left(\mathbf{E}_i^g \mid \mathbf{E}_i^q\right) - \frac{1}{|\mathcal{I}|} \sum_{j=1}^{|\mathcal{I}|} \log q_\phi\left(\mathbf{E}_j^g \mid \mathbf{E}_i^q\right)\right],
\end{aligned}
\end{equation}
where,~$\mathcal{L}_{m\_\text{dis}}$ is distancing loss. We highlight that the parameters of the network $q_\phi(\cdot)$, as well as $\mathbf{E}^g$ and $\mathbf{E}^q$, iterative updates. Up to this point, we can learn to get modal-unique by Eq.~\eqref{dis_mi}.

Although a variational distribution is used, the upper bound can still be maintained. Specifically, following~\cite{club}, we give the following theoretical analysis, and complete proof in~\ref{proof2}.
{\thm{Denote \( q_\theta(\mathbf{E}^g, \mathbf{E}^q) = q_\theta(\mathbf{E}^q \mid \mathbf{E}^g) p(\mathbf{E}^g) \). If
\[KL\left(p(\mathbf{E}^g, \mathbf{E}^q) \| q_\theta(\mathbf{E}^g, \mathbf{E}^q)\right) \leq KL\left(p(\mathbf{E}^g) p(\mathbf{E}^q) \| q_\theta(\mathbf{E}^g, \mathbf{E}^q)\right), then
\]
\(\mathbb{I}(\mathbf{E}^g ; \mathbf{E}^q) \leq \mathbb{I}_{\mathrm{upper}}(\mathbf{E}^g ; \mathbf{E}^q)\). The equality holds when \(\mathbf{E}^g\) and \(\mathbf{E}^q\) are independent.
}}
{\remark{After the theorem 2, we guarantee that even if we use the variational distribution, it remains as a mutual information upper bound. It also shows that we can theoretically learn modal-unique, which is more efficient and rigorous than orthogonal constraints.}}

\subsubsection{Maximizing the Lower Bound for Modal-generic Learning}
Cross-modal alignment models the correlation between modalities and allows the modalities to convey general information to each other~\cite{spc-common4}. 
We first model visual and text features into the same representation space, enabling interoperability between natural language and visual semantics. Previous works~\citeN{MMSSL, BM3} use InfoNCE~\cite{infonce} to obtain modal-generic representation. However, we argue that this strategy is time-inefficient. Thus we design a modalities alignment loss, SoloSimLoss, which is characterized by time-efficient and adaptively extracts modal-generic information from various modalities. The goal of this loss function~$\mathcal{L}_{\text{SoloSimLoss}}$ is to maximize the lower bound of MI for better alignment in visual-text pairs.
\begin{equation}
\mathcal{L}_{\text{SoloSimLoss}}=-\frac{1}{|\mathcal{I}|} \sum_{i=1}^{|\mathcal{I}|} \log (\textrm{Softmax} \{ \exp (\mathbf{E}_g^v \cdot \mathbf{E}_g^t / \tau) \}).
\end{equation}
Specifically, first, the inner product of the visual~$\mathbf{E}_g^v$ and textual~$\mathbf{E}_g^t$ representation of each sample is calculated. Then, a softmax function is applied to the results to obtain the similarity of each sample to all other samples. In addition, we define a temperature parameter~$\tau$ that acts as a scaling factor that adjusts the smoothness of the softmax distribution. We also name this loss function $\mathcal{L}_{\text{align}}$ for convenience in the later section.


\subsection{Fusion and Optimization}
We have obtained the representation of each modality of users and items under the heterogeneous graph, and the representation of the item obtained under the homogeneous graph, and modal-unique and learned modal-unique and modal-generic representation of the items. Now we conduct modal fusion to get the final representation of users~$\textbf{E}_u$ and items~$\textbf{E}_i$.
\begin{equation}
    \textbf{E}_i= \{\textbf{E}_q^t || \textbf{E}_g^t || \textbf{E}_q^v || \textbf{E}_g^v + \textbf{H}\}; \quad     \textbf{E}_u = \{\textbf{W}_t \cdot \textbf{E}_u^t || \textbf{W}_v \cdot \textbf{E}_u^v\}.
\end{equation}
So far, we have the final representation. Then we employ BPR~(Bayesian Personalized Ranking) as a fundamental part of our optimization. Specifically, in BPR, we denote the set of training triplets as $\mathcal{D}$, where each triplet $(u, i, j)$ represents a user $u$ preferring interacted item $i$ over uninteracted item $j$. The BPR loss function can be formulated as follows:
\begin{equation}
\mathcal{L}_{bpr}=-\sum_{(u, i, j) \in \mathcal{D}} \log \sigma\left(\mathbf{E}_u \cdot \mathbf{E}_i-\mathbf{E}_u \cdot \mathbf{E}_j \right),
\end{equation}
where $\sigma(x)$ denotes the sigmoid function.
We then combine the previous alignment loss and distancing loss~(Note that from Figure~\ref{Overall} we can see that we distance+ from the visual and textual modalities separately, hence there are two losses) to form the final loss function:
\begin{equation}
\mathcal{L}=\mathcal{L}_{\text {bpr}}+\alpha \cdot \mathcal{L}_{\text {align}}+ \beta \cdot (\mathcal{L}_{\text {v\_dis}} + \mathcal{L}_{\text {t\_dis}}),
\end{equation}
where $\alpha$ and $\beta$ are important hyperparameters and we will perform sensitivity analyses in the experimental section.

\section{Experiments}
In this section, we conduct various experiments to verify the effectiveness of our framework and various analytical experiments that demonstrate our generalization, flexibility, and efficiency. The entire experimental section is developed with the following questions~(RQs):
\begin{itemize}
  \item \textbf{RQ1: }Does our~\shortname~remain advanced in comparison to recently published methods?
  \item \textbf{RQ2: }Does each component of our~\shortname~benefit performance? 
  \item \textbf{RQ3: }How does the parameter tuning of~\shortname~affect model performance? 
  \item \textbf{RQ4: }How about a comprehensive analysis of our model in terms of generalization, flexibility, and efficiency?
  \item \textbf{RQ5: }Does analyzing~\shortname~visually demonstrate advantages?
\end{itemize}
\subsection{Experimental Settings}
\vspace{3pt}
\noindent
\textbf{A \text{-} Datasets.} Follow previous work~\citeN{freedom, LATTICE, mentor}, we conduct empirical study on three popular datasets. Specifically, we use \textbf{Baby}, \textbf{Sports}, and \textbf{Clothing}. All datasets are processed according to ~\textsf{MMRec}~\footnote{https://github.com/enoche/MMRec}~\cite{mmrec, dataset}. Table~\ref{tab:dataset_statistics} shows the basic information of the datasets.

\begin{table}[!ht]
\vskip -0.1in
    \centering
\caption{Statistics of three popular datasets.}
\vskip -0.05in
\label{tab:dataset_statistics}
    \begin{tabular}{ccrrr}
    \toprule
         \textbf{Dataset}&  \# \textbf{Users}&  \# \textbf{Items}&  \# \textbf{Interaction}& \textbf{Sparsity}\\
         \midrule
         Baby &  19,445 &  7,050 &  160,792 & 99.88\%\\
         Sports &  35,598 &  18,357 &  296,337 & 99.95\%\\
         Clothing &  39,387 &  23,033 &  278,677 & 99.97\%\\
         \bottomrule
    \end{tabular}
\vskip -0.15in
\end{table}

\vspace{5pt}
\noindent
\textbf{B \text{-} Evaluation Protocols.}
Following the usual principle~\citeN{freedom, LATTICE, mentor} of a fair comparison, we use the two Top\text{-}\textit{K} metrics, Recall~(R@\textit{k}) and NDCG@K~(N@\textit{k})~\cite{shisong1, shisong2} to measure performance. Specifically, we set $k$ to 10 and 20. The dataset is divided according to the popular partitioning criterion~\citeN{mentor} of 8:1:1 for training: validation: testing. Limited by length, the comparison method is at~\ref{compared}.


\vspace{5pt}
\noindent
\textbf{C \text{-} Implementation details.}
To verify the effectiveness of our framework more fairly and open source, we implement our framework using the popular benchmark~\textsf{MMRec}~\cite{mmrec}. We observe that the popular baseline results remain consistent in~\cite{freedom, mentor, mmrec_survey}, both of them are based on~\textsf{MMRec}. Therefore, in order to maintain a fair performance comparison, for the baseline results, we use the results of the original~\cite{freedom, mentor} as the results of our baseline.  For our framework, consistent with all baselines, we initialize the embedding with Xavier initialization\cite{Xavier} of dimension 64 and optimize~\shortname with Adam optimizer. The learning rate is fixed at 1e-4, and the number of GCN layers in two types of graph is set to $2$. 

\begin{table*}[!ht]
    \centering
    \caption{Performance comparison of baselines and our framework in terms of Recall@K (R@\textit{K}), and NDCG@K (N@\textit{K}).}
    \tabcolsep=0.1cm
    \vskip -0.1in
    \label{tab:comparison results}
    \begin{tabular}{r|cccc|cccc|cccc}
    \toprule[1.2pt]
         \textbf{Datasets}&  \multicolumn{4}{c}{\textbf{Baby}}&  \multicolumn{4}{|c}{\textbf{Sports}}&  \multicolumn{4}{|c}{\textbf{Clothing}}\\
         \midrule
         \midrule
         \textbf{Methods} & \textbf{R@10} & \textbf{R@20} & \textbf{N@10} & \textbf{N@20} & \textbf{R@10} & \textbf{R@20} & \textbf{N@10} & \textbf{N@20} & \textbf{R@10} & \textbf{R@20} & \textbf{N@10} & \textbf{N@20} \\ 
         \midrule
         MF-BPR~(UAI'09) & 0.0357& 0.0575& 0.0192& 0.0249& 0.0432& 0.0653& 0.0241& 0.0298 & 0.0187 & 0.0279 & 0.0103 & 0.0126\\
         LightGCN~(SIGIR'20) & 0.0479& 0.0754& 0.0257& 0.0328& 0.0569& 0.0864& 0.0311& 0.0387 & 0.0340 & 0.0526 & 0.0188 & 0.0236\\
         LayerGCN~(ICDE'23) & 0.0529& 0.0820& 0.0281& 0.0355& 0.0594& 0.0916& 0.0323& 0.0406& 0.0371& 0.0566& 0.0200& 0.0247\\
         \midrule
         VBPR~(AAAI'16) &  0.0423&  0.0663&  0.0223&  0.0284&  0.0558&   0.0856& 0.0307& 0.0384& 0.0281& 0.0415& 0.0158& 0.0192\\
         MMGCN~(MM'19) &  0.0378&  0.0615&  0.0200&  0.0261&  0.0370&   0.0605& 0.0193& 0.0254 & 0.0218& 0.0345& 0.0110& 0.0142\\
         DualGNN~(TMM'21) &  0.0448&  0.0716&  0.0240&  0.0309&  0.0568&   0.0859& 0.0310& 0.0385& 0.0454& 0.0683& 0.0241& 0.0299\\
         LATTICE~(MM'21) &  0.0547&  0.0850&  0.0292&  0.0370&  0.0620&  0.0953&  0.0335&  0.0421& 0.0492& 0.0733& 0.0268& 0.0330\\
          SLMRec~(TMM'22) & 0.0529& 0.0775& 0.0290& 0.0353& 0.0663&   0.0990& 0.0365& 0.0450 &0.0452& 0.0675& 0.0247& 0.0303\\
          MICRO~(TKDE'22) & 0.0584& 0.0929& 0.0318& 0.0407& 0.0679&   0.1050& 0.0367& 0.0463 &0.0521& 0.0772& 0.0283& 0.0347\\
          PAMD~(WWW'22) & 0.0618 & 0.0957 & 0.0334 & 0.0422 & 0.0723 & 0.1101 & 0.0393 & 0.0491 & 0.0634 & 0.0942 & 0.0344 & 0.0421\\
         BM3~(WWW'23) & 0.0564& 0.0883& 0.0301& 0.0383& 0.0656& 0.0980& 0.0355& 0.0438& 0.0422& 0.0621& 0.0231& 0.0281\\
         MMSSL~(WWW'23) & 0.0613& 0.0971& 0.0326& 0.0420& 0.0673& 0.1013& 0.0380& 0.0474& 0.0531& 0.0797& 0.0291& 0.0359\\
          FREEDOM~(MM'23) & 0.0627& 0.0992& 0.0330& 0.0424& 0.0717& 0.1089& 0.0385& 0.0481& 0.0629& 0.0941& 0.0341& 0.0420\\
          MGCN~(MM'23) & 0.0620& 0.0964& 0.0339& 0.0427& \underline{0.0729}& \underline{0.1106}& \underline{0.0397} & \underline{0.0496}& \underline{0.0641}& \underline{0.0945}& \underline{0.0347}& \underline{0.0428}\\
         LGMRec~(AAAI'24) & \underline{0.0644}& \underline{0.1002}& \underline{0.0349}& \underline{0.0440}& 0.0720& 0.1068& 0.0390& 0.0480 & 0.0555& 0.0828& 0.0302& 0.0371\\
         \rowcolor{black!15} \shortname~(Ours) &  \textbf{0.0663}&  \textbf{0.1033}& \textbf{0.0354}&  \textbf{0.0449}&  \textbf{0.0757}&  \textbf{0.1145} &\textbf{0.0408} & \textbf{0.0508} 
         &  \textbf{0.0659}&  \textbf{0.0953} &\textbf{0.0358} & \textbf{0.0432}  \\
         \bottomrule[1.2pt]
    \end{tabular}
\end{table*}

\subsection{Performance Comparison~(RQ1)}
To verify the superiority of our proposed framework, we compare~\shortname with the kinds of existing well-known methods. We present the results on three popular datasets as well as four commonly used metrics. Table~\ref{tab:comparison results} shows the results of the experiment and we obtain the following observations:

\vspace{3pt}
\noindent
\textbf{Obs~1. Our method is consistently better than all baselines across all metrics.} Specifically, on the sports dataset, we improve 3.84\% (R@10) and 7.21\% (R@20) compared to the best baseline. We attribute the enhancement that we split the initial representation and perform double bounds of mutual information to obtain a more expressive representation. With this strategy, we can make the overall representation of each modality more complete. Most baselines either just utilize more graph structure or just align modality using self-supervised learning, ignoring modal-unique information. Although some methods~\citeN{MICRO, MGCN, PAMD} attempt to disentangle the modal-generic and -unique, they simply split the modal representation into two parts by subtraction, and do not constrain both by explicitly aligning the modal-generic representation and distancing modal-unique representation, hence the performance potential is not better exploited.

\vspace{3pt}
\noindent
\textbf{Obs~2. The method that includes obtaining modal-unique is better than most alignment-only methods.} In terms of overall performance, methods that distinguish between modal-unique and -generic representation (e.g. MICRO, PAMD, MGCN) are generally superior to traditional GNN or SSL-based methods that don't distinguish both, such as MMGCN, MMSSL, and BM3. These methods only align all the information between different modalities so that they are in the same representation space, ignoring the specific effects of modal-unique information, resulting in suboptimal performance. This finding also validates the point of some of the methods that modality is good, but it should not be the entirety. It is important and necessary for multimedia recommendation methods to distinguish between modality-unique and -generic representations.


\subsection{Ablation Study~(RQ2)}
Previously, we observed the overall performance of~\shortname. Now, we take it a step further and explore the performance impact of each module of~\shortname~with ablation experiments. We design three variants:
\begin{itemize}
  \item \textbf{w/o All}: We remove both distancing and alignment modules.
  \item \textbf{w/o Distancing}: We remove the representation distancing module to eliminate modal-unique effects.
  \item \textbf{w/o Alignment}: We remove the representation alignment module to eliminate modal-generic effects.
  \item \textbf{\shortname}: We retain all modules for comprehensive comparison.
\end{itemize}
\begin{figure}[!ht]
	\centering
    {\includegraphics[width=.49\linewidth]{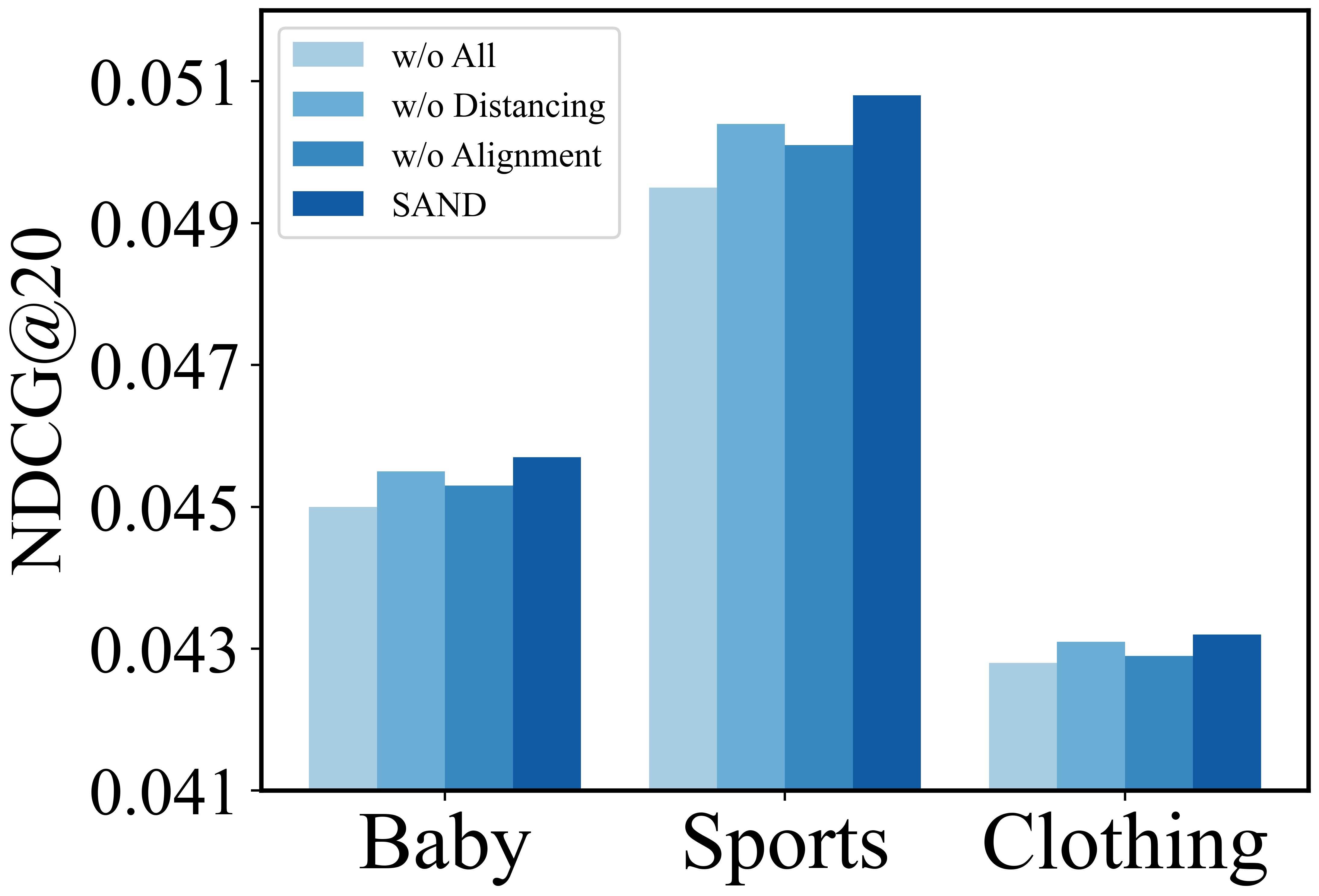}}
    {\includegraphics[width=.49\linewidth]{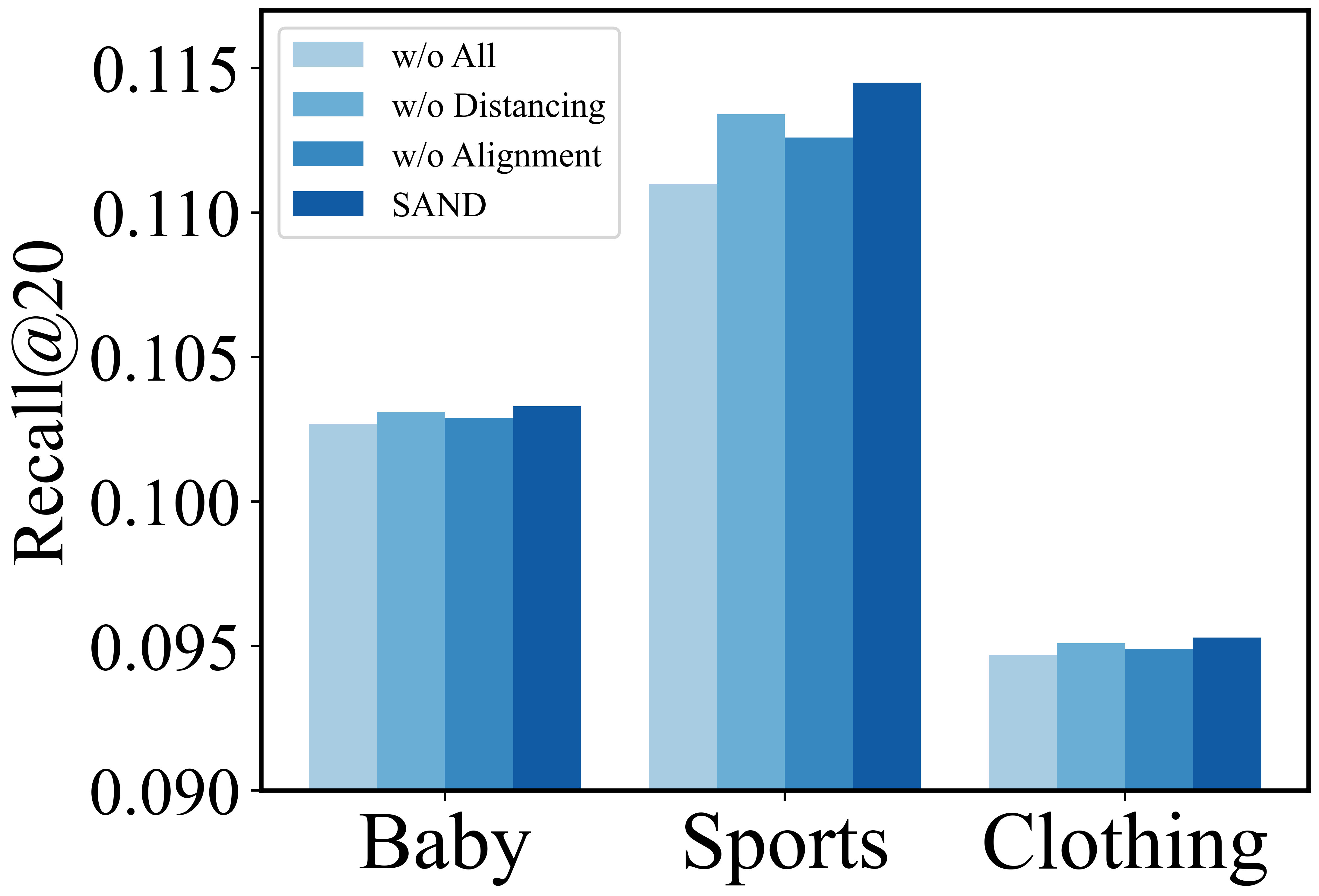}}
   \caption{The effect of each module on~\shortname.}
\vspace{-2mm}
\label{ablation}
\end{figure}
Specifically, as shown in Figure~\ref{ablation}, we obtain the following observations in three datasets:

\vspace{3pt}
\noindent
\textbf{Obs~3. Each module has a beneficial effect.} On all three datasets, we find that adding any of our proposed modules to the model is able to improve the results, compared to the case where no module is added.
This is partly due to the success of modal alignment, which models the representation of different modal patterns in the same space, allowing for more subtle modal fusion. On the other hand, the preservation of modal-unique gives the models a relatively good performance improvement.

\vspace{3pt}
\noindent
\textbf{Obs~4. Multiple modules combine to yield better performance.} 
We observe that the combination of modules produces the best results and consider that this matches our motivation. This is due to our separate learning of modal-unique and modal-generic features, which complement each other with a more powerful representation. In addition, the performance improvements observed through module combinations highlight the collaborative effects of module integration. \shortname can further unleash its performance potential through more reasonable constraints to delineate the modal-generic and -unique parts.

\begin{figure}[t]
    \centering
    {
        \includegraphics[width=0.323\linewidth]{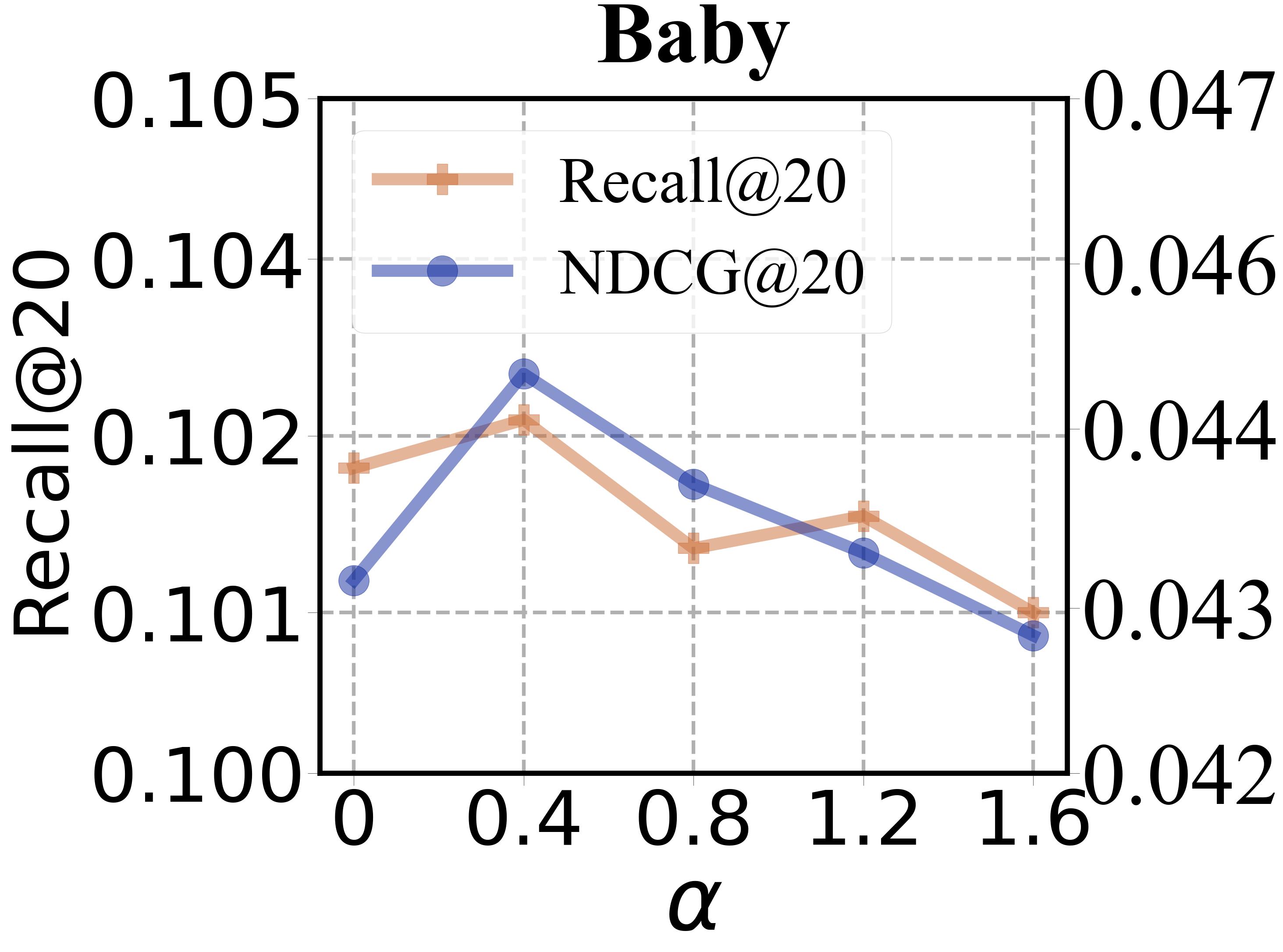}
    }
    {
        \includegraphics[width=0.303\linewidth]{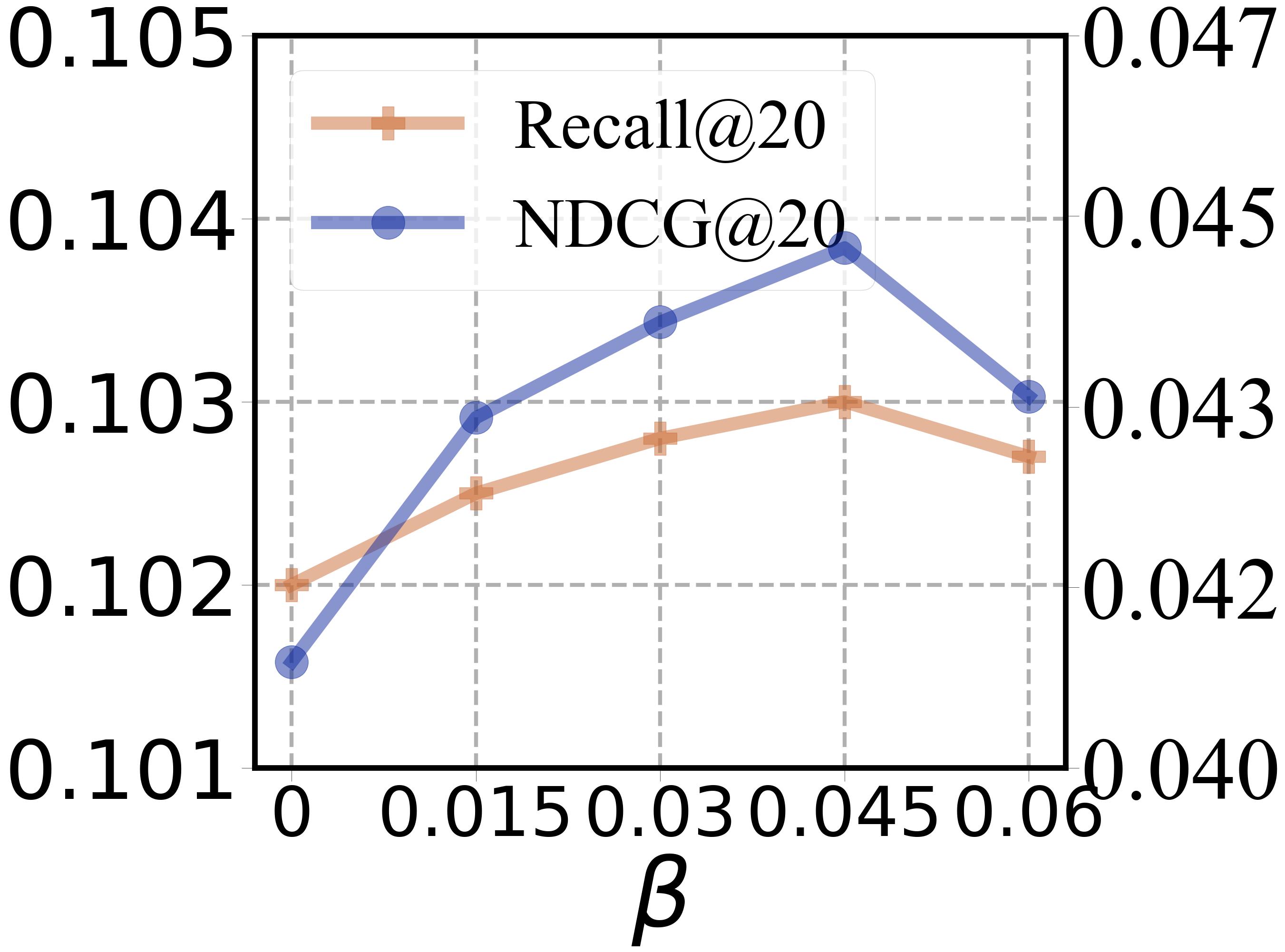}
    }
    {
        \includegraphics[width=0.323\linewidth]
        {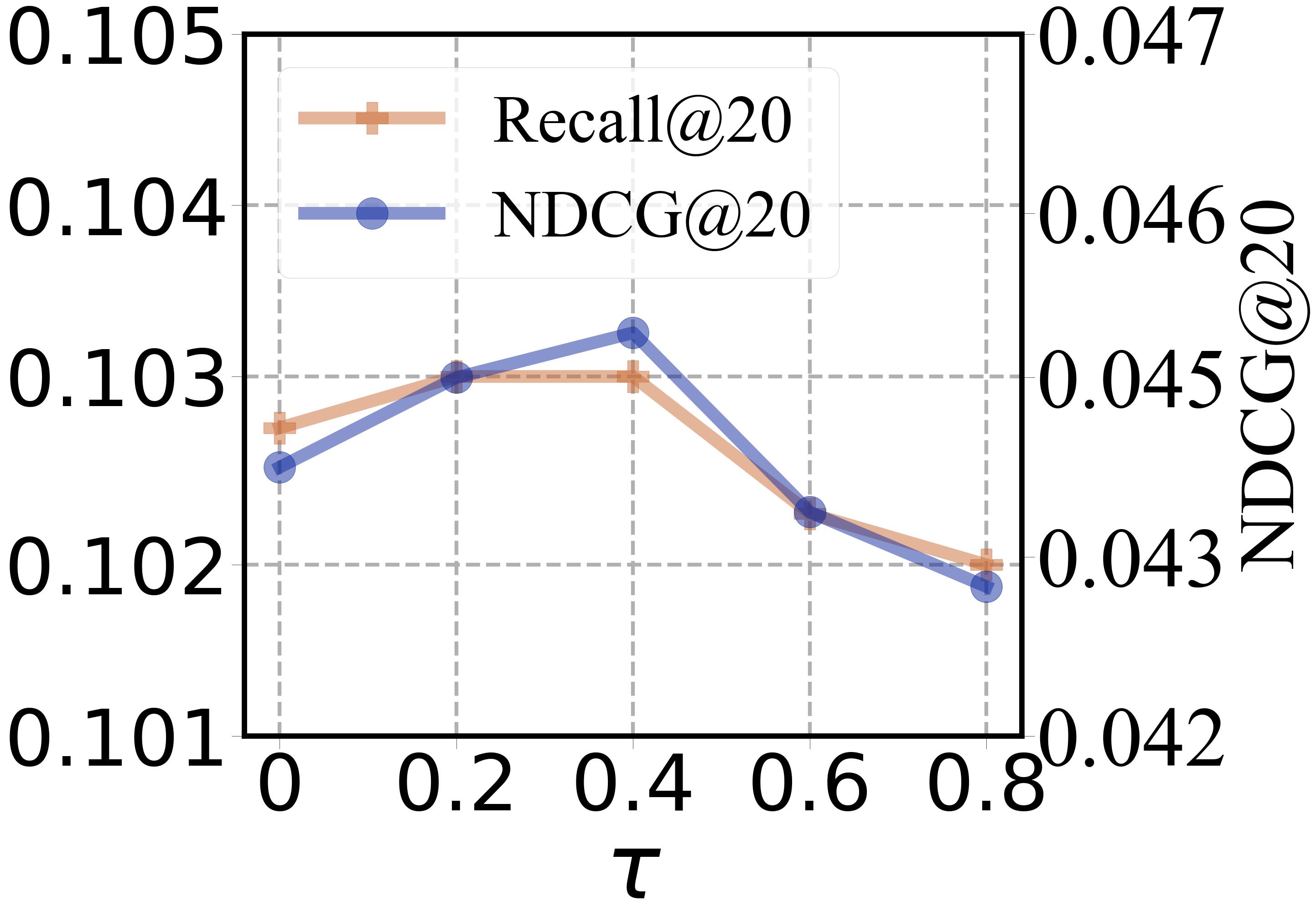}
    }\\
    {
        \includegraphics[width=0.322\linewidth]{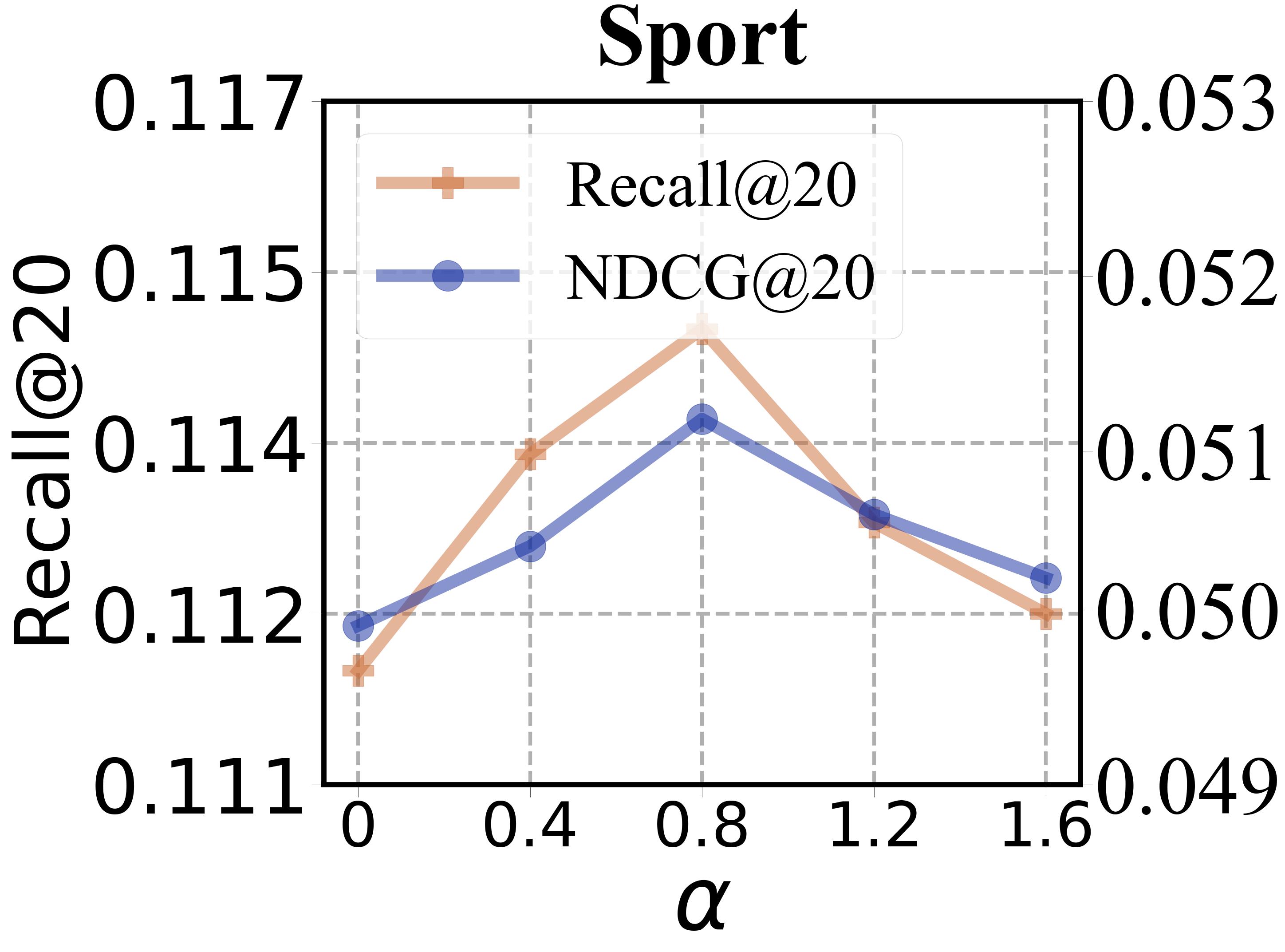}
    }
    {
        \includegraphics[width=0.302\linewidth]{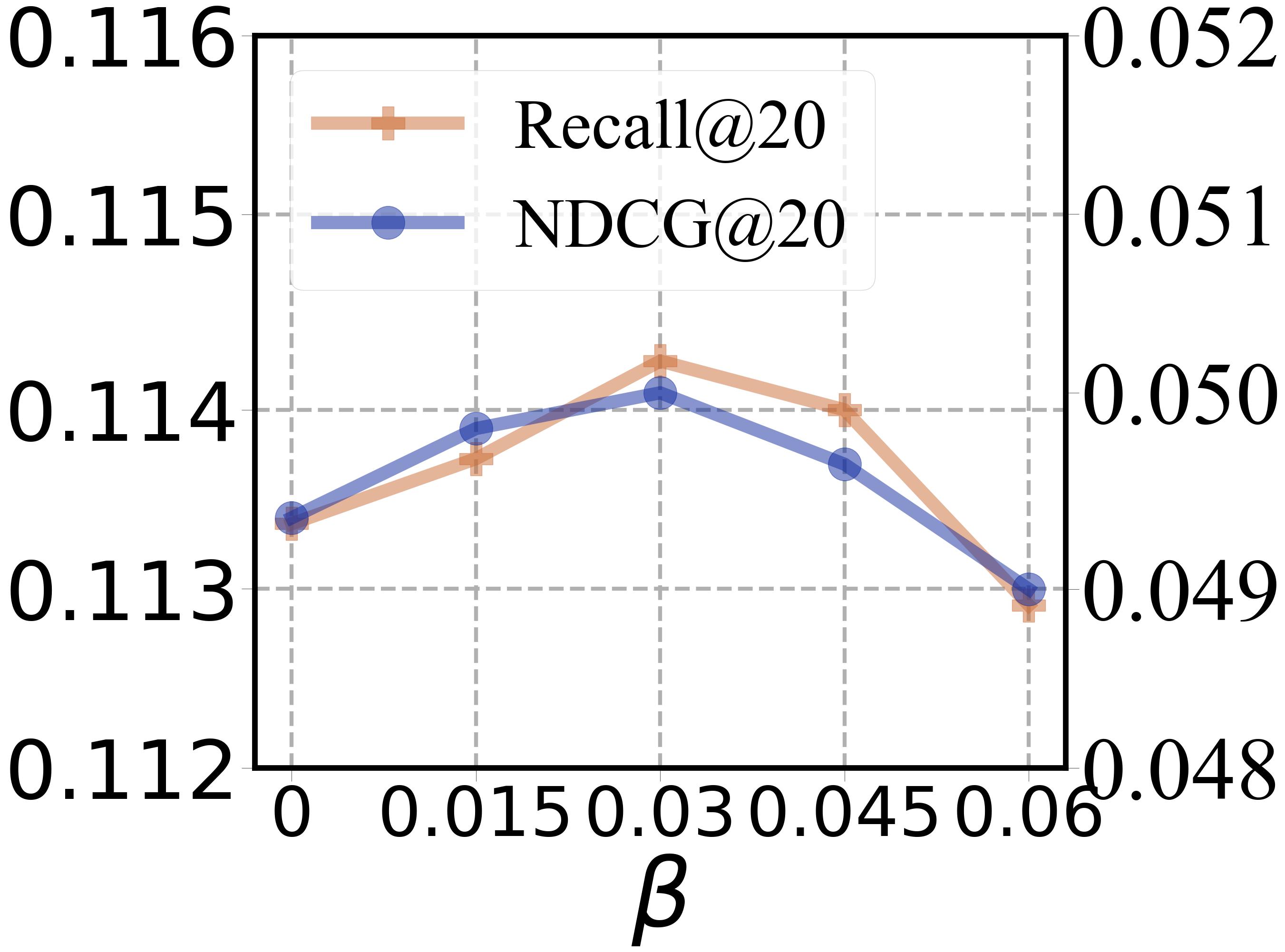}
    }
    {
        \includegraphics[width=0.322\linewidth]{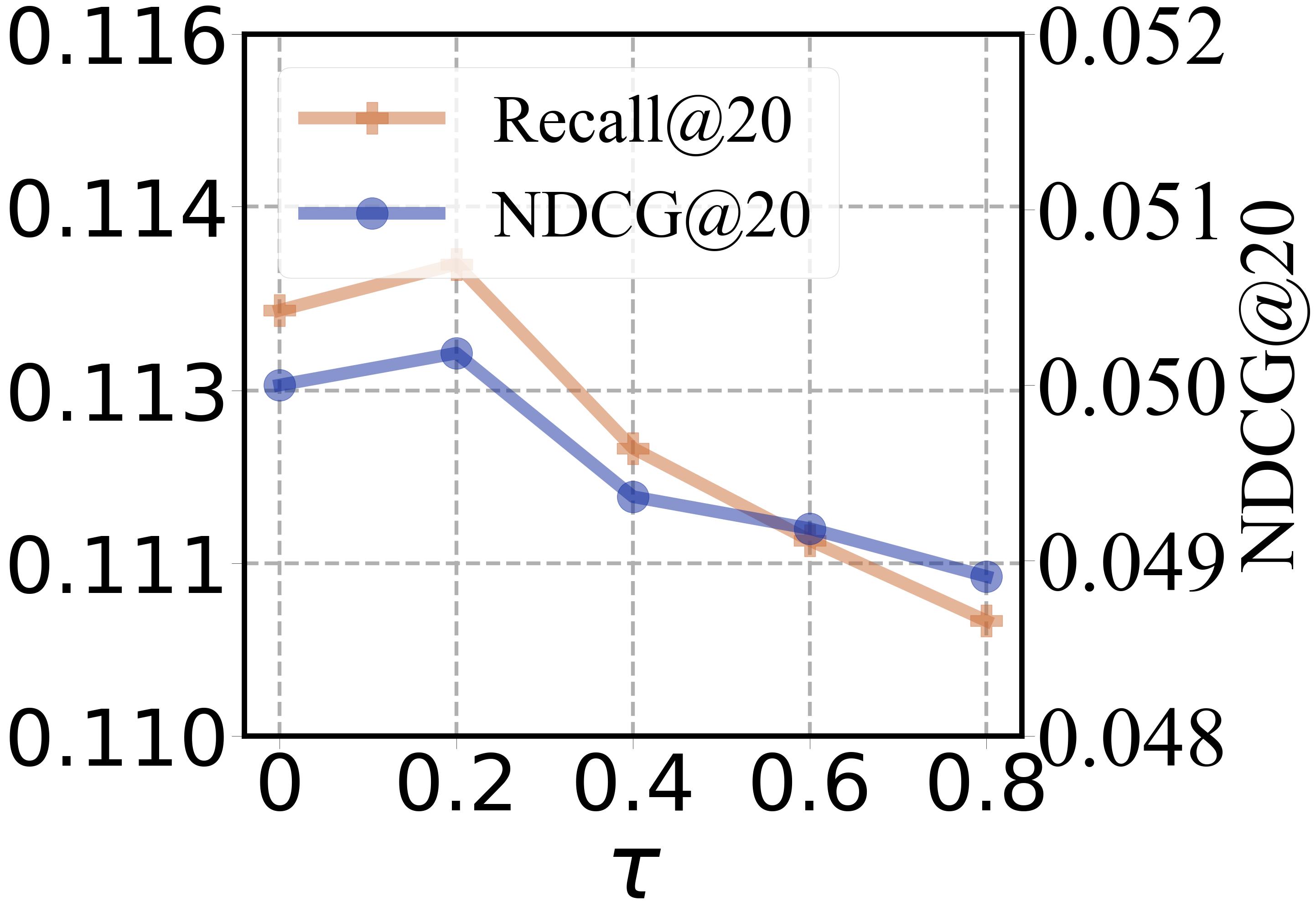}
    }
    \vspace{-5pt}
    \caption{Impact of the alignment weight $\alpha$, distancing weight $\beta$ and temperature coefficient $\tau$.}
    \label{PSA}    
    \vspace{-5pt}
\end{figure}

    

\subsection{Parameter Sensitivity Analysis~(RQ3)}
In this section, we experimentally explore the sensitivity under different parameter settings to understand performance trends. Our framework involves three main parameters,~\text{i}) the weight~$\alpha$ of the alignment module,~\text{ii}) the weight~$\beta$ of the distancing module, and~\text{iii}) the temperature coefficient~$\tau$. The experiment is conducted as a controlled variable in the Baby and Sports dataset. As shown in Figure~\ref{PSA}, we can observe that:

\vspace{3pt}
\noindent
\textbf{Obs~5. As the value of the parameter increases, there is a tendency for the performance to increase and then decrease.} As the values of the parameters increase, the performance of the model may improve as well, as an increase in the values of the parameters may enhance the ability of the model to fit the data better. For example, as $\alpha$ increases, it will pull the representation closer together well to make them in the same space, allowing for a finer modal fusion. However, as it continues to increase, it pulls the information too close together, resulting in impaired information on both sides, which in turn leads to a decrease in performance. The other two parameters are similarly so.

\subsection{Framework Generalization, Flexibility, Efficiency Analysis~(RQ4)}
In this section, we perform a comprehensive analysis, in terms of generalization, flexibility, and efficiency, respectively. To demonstrate the generalization of our framework, which allows us to plus the alignment and distancing modules on other methods; as well as the flexibility of our framework itself, which allows us to replace the modules in it to adapt to various available resource conditions; and the efficiency of our framework, which compares the time and memory under different modules.

\vspace{5pt}
\noindent
\textbf{A \text{-} Generalization Analysis.}
Our alignment and distancing modules are not only applicable to our proposed GNN-based Encoder. It can also be applied to other models. We select two popular methods plus our two modal strategies. The VBPR multimodal model and the GNN-based FREEDOM. The results are shown in Table~\ref{Generalization Analysis}. We observe that both VBPR and FREEDOM have different performance improvements after incorporating our alignment and distancing modules, and FREEDOM's performance improvement is slightly lower than VBPR's due to its own powerful multi-graph convolutional structure.
We get the following Conclusion:

\vspace{2pt}
\noindent
\textbf{Con~1. Our framework is a \underline{plug-and-play} module that can be easily integrated into various multimedia recommendation models and achieve better results.} 

\begin{table}[!h]
\vskip -0.01in
    \centering
    \caption{Generalization analysis on Baby dataset.}
    \vskip -0.12in
    \label{Generalization Analysis}
    \begin{tabular}{lcrrr}
    \toprule
         \quad \quad  \textbf{Model}& \textbf{R@10}& \textbf{R@20}&  \textbf{N@10}& \textbf{N@20}\\
         \midrule
         \quad \quad VBPR &  0.0423 &  0.0663&  0.0223&  0.0284\\
         VBPR + \shortname &  0.0457 &  0.0743 &  0.0243 & 0.0317\\
         \quad \quad Improv. & 8.04\% & 12.07\% & 8.97\% & 11.62\%  \\
         \midrule
         \quad \quad FREEDOM &  0.0627& 0.0992& 0.0330& 0.0424\\
         FREEDOM + \shortname &  0.0636 &  0.1013 &  0.0336 & 0.0433\\
         \quad \quad Improv. & 1.44\% & 2.12\% & 1.82\% & 2.12\% \\
         \bottomrule
    \end{tabular}
\vskip -0.15in
\end{table}

\vspace{5pt}
\noindent
\textbf{B \text{-} Flexibility Analysis.} We provide \underline{2x2} combinations of four methods to extract modal-generic and -unique representation. In particular, mutual information minimization (MIM) is proposed for the distancing module. We additionally give negative $\ell_2$ loss~($\ell_2$) in~\ref{l2} to compare it. Solosimloss and InfoNCE loss are designed for the alignment module. Therefore many modules can be replaced, and here we experiment with various combinations of modules. The results of various combinations of experiments are shown in Table~\ref{Flexibility Analysis}. We find that Solosimloss and MIM achieve the best results. \shortname$_a$ performs slightly better than \shortname$_c$, which implies that there may be a large number of pseudo-negative samples between modalities (e.g. textual descriptions that are similar to the anchor image). Therefore, aligning only inter-modal positive samples may be sufficient. In addition, even the worst-performing combination, \shortname$_d$, can compete with the optimal baseline, LGMRec.

\vspace{2pt}
\noindent
\textbf{Con~2. Our framework is \underline{effective} and \underline{flexible}, providing four combinations to better delineate between modal-generic and -unique representation to enhance items representation and achieve optimal or sub-optimal results.} 


\begin{table}[!h]
\vskip -0.02in
    \centering
\caption{Flexibility analysis on Baby dataset.}
\vskip -0.12in
\small
\label{Flexibility Analysis}
\setlength\tabcolsep{2.1pt}{
    \begin{tabular}{cccrrrrrr}
        \toprule
        \textbf{Variant} & \textbf{Solosim} & \textbf{InfoNCE} & $\mathrm{\textbf{MIM}}$ & \textbf{$\ell_2$} & \textbf{R@10} & \textbf{R@20} & \textbf{N@10} & \textbf{N@20} \\
        \midrule
        \shortname$_a$ & $\checkmark$ &   & $\checkmark$  &  &  \textbf{0.0663} & \textbf{0.1033} &\textbf{0.0354}& \textbf{0.0449}\\
        \shortname$_b$ & $\checkmark$ &   &   & $\checkmark$  &  0.0650 &  0.1012 &  0.0349 &  0.0441\\
        \shortname$_c$ &  &  $\checkmark$ &  $\checkmark$ &   & 0.0657 &  0.1026 &  0.0348 &  \textbf{0.0449} \\
        \shortname$_d$ &  &  $\checkmark$ &   & $\checkmark$  & 0.0644 &  0.1007 &  0.0345&  0.0436\\
        \bottomrule
    \end{tabular}}
    \vskip -0.15in
\end{table}

\vspace{5pt}
\noindent
\textbf{C \text{-} Efficiency Analysis.} In this section, we compare the efficiency of each model. We conduct an efficiency analysis of several variants mentioned in the previous flexibility analysis.
We perform the following combinations to validate efficiency.
Specific statistical memory and time are shown in Table~\ref{effiency}. We observe an increase in both memory and training time for the multimodal model. This is because this adds more modal features compared to the generic CF model. With several of our methods, it is evident that the $\mathrm{MIM}$-based approach has a large increase in resource consumption. We readily acknowledge that this is indeed a limitation of our framework. However, it also points to a direction for future research, where we aim to develop methods that consume fewer resources while offering better performance.


\subsection{Visual Validation of Motivation~(RQ5)}
\begin{figure}
 	\centering
 \includegraphics[width=0.45\textwidth]{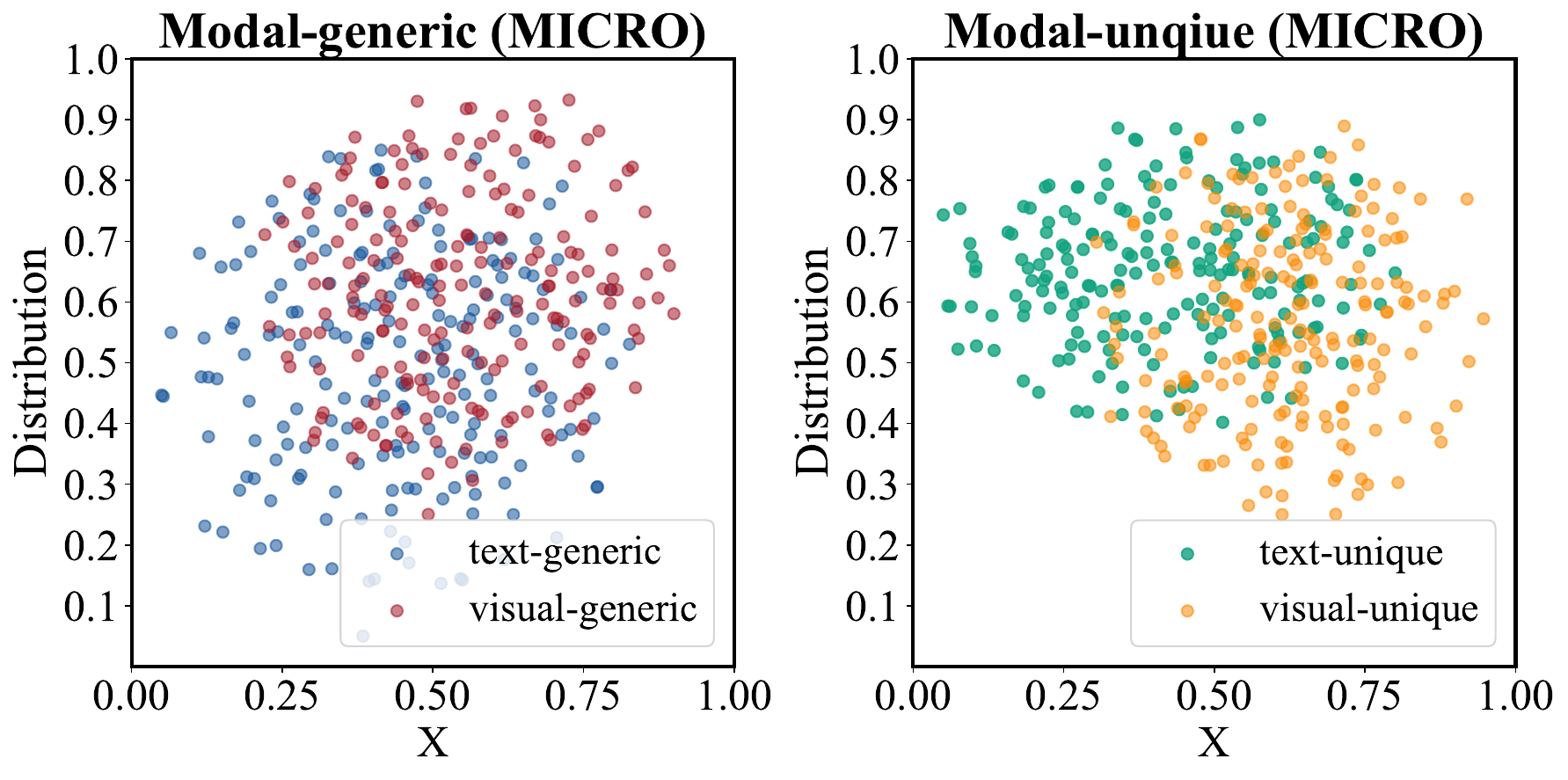} 
 \vspace{-3mm}
   \caption{Distribution of representation obtained by MICRO with modal-generic part on the left and modal-unique part on the right.}
 \vspace{-3mm}
\label{visual1}   
\end{figure}

\begin{figure}
 	\centering
 \includegraphics[width=0.45\textwidth]{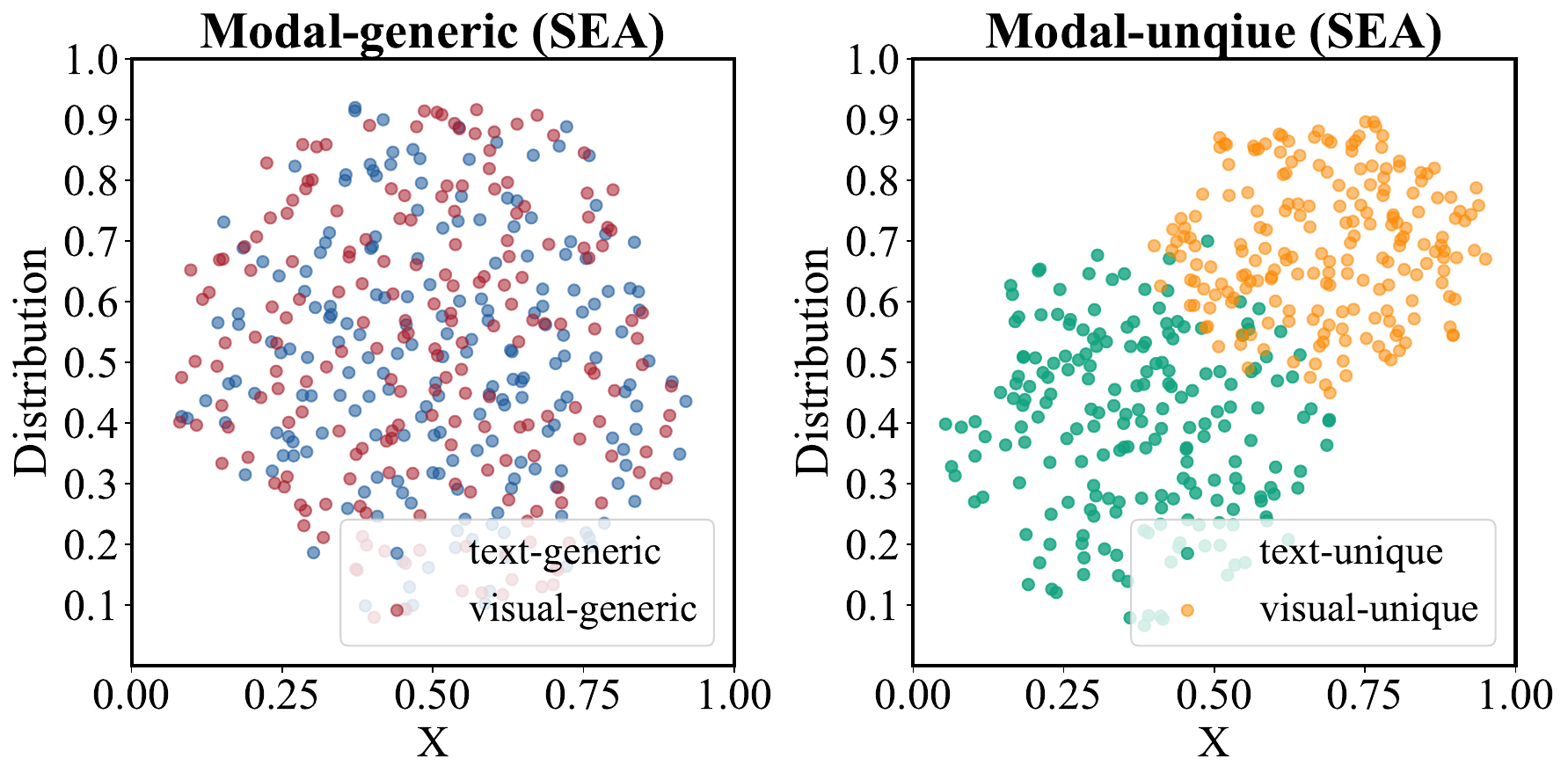} 
 \vspace{-3mm}
   \caption{Distribution of representation obtained by \shortname  with modal-generic part on the left and modal-unique part on the right.}
    \vspace{-3mm}
\label{visual2}   
\end{figure}

To gain a better understanding of the advantages of our modal representation, we take a step: randomly pick 200 items from the Baby dataset and use the TSNE~\cite{tsne} algorithm to map MICRO~(i.e. Learning modal-unique using orthogonal constraint) and \shortname~ representation into a two-dimensional space. 
Upon analyzing the distribution of 2D features from Figure~\ref{visual1} and Figure~\ref{visual2}, we observe that the distributions of the MICRO representations are still partially coupled, which validates our previous motivation and theoretical analysis that a modal-unique and -generic representation of distinguishability cannot really be obtained using only subtraction and orthogonality constraints, whereas \shortname~'s modal-generic representation tends to align for visual and textual modalities, while modal-generic and -unique representations are well-distanced across modalities. Meanwhile, text and visual modal-unique parts seem to capture a different distribution of information. \shortname~only aligns modal-generic parts while retaining modal-unique parts to serve the final recommendation, avoids the performance loss caused by aligning unnecessary information between modalities between modalities, and ensures state-of-the-art performance. Finally, we give the last conclusion: 

\vspace{2pt}
\noindent
\textbf{Con~3. Compared to using orthogonal constraints (subtraction) methods, our proposed method establishes more rigorous double bounds of MI to achieve higher quality modal-generic and -unique representations, resulting in better effectiveness and generalizability. }


\section{Related Works}
Multimedia recommendation has been widely studied because it introduces a rich variety of modal information that enhances the effectiveness of recommendation. We categorise existing methods into two groups, GNN-based~\citeN{MMGCN,LATTICE,dualgnn,freedom,guo2023lgmrec} and SSL-based~\citeN{SLMRec, MICRO, LATTICE, MMSSL,BM3,MGCN,mentor,alignrec}, based on their technical characteristics.

\vspace{3pt}
\noindent
\textbf{GNN\textbf{-}based Methods.} The impressive representational modelling capabilities of graph neural networks have led to significant improvements not only in collaborative filtering methods (e.g., LightGCN), but also in multimedia recommendation~\citeN{MMGCN,LATTICE,dualgnn,freedom, guo2023lgmrec}. MMGCN~\cite{MMGCN} firstly uses GCN to model two interaction graphs, containing visual and textual information respectively, and then fuses the item representation obtained from each party to obtain the final representation. LATTICE~\cite{LATTICE} argue that the underlying structural information behind the modalities reflects the underlying relationships between the modalities, and therefore not only construct interaction graphs as well as item homogeneity graphs learnt from the modal features, inference yields more expressive representation. DualGNN~\cite{dualgnn} differs from LATTICE which focuses on obtaining the user-user co-occurrence graph and enhancing the users' modal representation. LGMRec~\cite{guo2023lgmrec} utilises hypergraph neural networks for capturing higher-order interactions, and richer modal fusion.

\vspace{3pt}
\noindent
\textbf{SSL\textbf{-}based Methods.} Self-supervised learning~(SSL) effectively mitigates supervised signal sparsity, advancing the fields of computer vision~\cite{ssl-nlp-survey}, natural language processing~\cite{ssl-nlp-survey}, recommendation~\cite{ssl-rec-survey}, and so on. The SSL-based multimedia recommendation~\citeN{SLMRec, MICRO, LATTICE, MMSSL, BM3, MGCN, mentor,alignrec} builds on the success of the GNN-based methods. It excels in extracting potential relationships between modalities~\cite{SLMRec}, modal feature fusion~\cite{BM3, MICRO}, and mitigating label sparsity~\cite{MMSSL}. For example, SLMRec~\cite{SLMRec} explores potential relationships between modalities using contrastive learning to produce powerful representation. BM3~\cite{BM3} builds multiple views by dropout strategy and reconstructs the interaction graph, and designs the intra- and inter-modal contrastive loss to learn the representation with promising results. MMSSL~\cite{MMSSL} mitigats label sparsity with two-stage self-supervised learning, generation and comparison, and achieved excellent performance.

\vspace{3pt}
\noindent
\textbf{Comparison with Our Method.} MICRO~\cite{MICRO}, PAMD~\cite{PAMD}, and MGCN~\cite{MGCN} first extract the unique parts of the modality, then use orthogonal constraints to learn modal-unique features. We completely abandon this approach. Instead, we split the representations and optimize the upper bound of the mutual information between the unique and general parts within the same modality to learn modal-unique features. This method is more reasonable and rigorous. Moreover, the most related work to ours is SimMMDG~\cite{SimMMDG}. However, it addresses domain generalization and modality missing issues, which are not feasible in multimedia recommendation. Our method is more flexible, with all modules being adaptable to complex situations.

\section{Conclusion}
In this paper, we propose a novel framework~\shortname~for multimedia recommendation. 
We begin by thoroughly analyzing the drawbacks of the existing orthogonal learning paradigm, which inspired the design of our proposed framework. Specifically, we first divide the representation of each modality into unique and generic parts. Then, we utilize a contrastive log-ratio upper bound to minimize the mutual information between the general and unique parts within the same modality. This increases the distance between their representations, thereby learning more discriminative modality-unique and generic features. Additionally, we design Solosim loss to maximize the mutual information of the generic parts across different modalities. By optimizing the lower bound of their mutual information, we align their representations, thus learning higher-quality modality-general features. Finally, extensive experiments on three datasets demonstrate the effectiveness, flexibility, and generalizability of our proposed framework.

\bibliographystyle{ACM-Reference-Format}
\balance
\bibliography{SAND}

\clearpage

\appendix
\section{APPENDIX}
\subsection{Proof of THEOREM 1}
\label{proof1}
Let \(\boldsymbol{x} = (x_1, x_2, \ldots, x_n)\) and \(\boldsymbol{y} = (1, 0, \ldots, 0)\) be unit vectors. We are interested in the distribution of the angle between the random vector \(\boldsymbol{x}\) and the fixed vector \(\boldsymbol{y}\). Due to isotropy, we only need to consider the randomness of \(\boldsymbol{x}\) as a unit vector while fixing \(\boldsymbol{y}\). Transforming \(\boldsymbol{x}\) into spherical coordinates, we get:
\begin{equation}
\left\{\begin{aligned}
x_1 &= \cos(\varphi_1) \\
x_2 &= \sin(\varphi_1) \cos(\varphi_2) \\
x_3 &= \sin(\varphi_1) \sin(\varphi_2) \cos(\varphi_3) \\
& \vdots \\
x_{n-1} &= \sin(\varphi_1) \cdots \sin(\varphi_{n-2}) \cos(\varphi_{n-1}) \\
x_n &= \sin(\varphi_1) \cdots \sin(\varphi_{n-2}) \sin(\varphi_{n-1})
\end{aligned}\right.
\end{equation}
where \(\varphi_{n-1} \in [0, 2\pi)\) and the remaining \(\varphi_i\) (for \(i < n-1\)) range from \([0, \pi]\). The angle between \(\boldsymbol{x}\) and \(\boldsymbol{y}\) is:
\begin{equation}
\arccos \langle \boldsymbol{x}, \boldsymbol{y} \rangle = \arccos x_1 = \arccos \cos(\varphi_1) = \varphi_1
\end{equation}
Thus, the angle \(\theta\) is represented by \(\varphi_1\).

To find the probability that the angle \(\theta\) is less than or equal to \(\theta\), we compute:
\begin{equation}
P_n(\varphi_1 \leq \theta) = \frac{\text{Integral of } \varphi_1 \text{ not exceeding } \theta \text{ on the spherical surface }}{\text{Total integral on the spherical surface }}
\end{equation}

The differential element of the integral on the spherical surface is \(\sin^{n-2}(\varphi_1) \sin^{n-3}(\varphi_2) \cdots \sin(\varphi_{n-2}) \, d\varphi_1 \, d\varphi_2 \cdots d\varphi_{n-1}\), thus:
\begin{equation}
\begin{aligned}
P_n(\varphi_1 \leq \theta) &= \frac{\int_0^{2 \pi} \cdots \int_0^\pi \int_0^\theta \sin^{n-2}(\varphi_1) \cdots \sin(\varphi_{n-2}) \, d\varphi_1 \, d\varphi_2 \cdots d\varphi_{n-1}}{\int_0^{2 \pi} \cdots \int_0^\pi \int_0^\pi \sin^{n-2}(\varphi_1) \cdots \sin(\varphi_{n-2}) \, d\varphi_1 \, d\varphi_2 \cdots d\varphi_{n-1}} \\
&= \frac{\text{Surface area of } (n-1)-dim \text{ sphere } \times \int_0^\theta \sin^{n-2} \varphi_1 \, d\varphi_1}{\text{Surface area of } n-dim \text{ sphere}} \\
&= \frac{\Gamma\left(\frac{n}{2}\right)}{\Gamma\left(\frac{n-1}{2}\right) \sqrt{\pi}} \int_0^\theta \sin^{n-2} \varphi_1 \, d\varphi_1
\end{aligned}
\end{equation}

This shows that the probability density function of \(\theta\) is:
\begin{equation}
p_n(\theta) = \frac{\Gamma\left(\frac{n}{2}\right)}{\Gamma\left(\frac{n-1}{2}\right) \sqrt{\pi}} \sin^{n-2} \theta
\end{equation}

When we are interested in the distribution of \(\eta = \cos \theta\), we need to perform a change of variables for the probability density function:
\begin{equation}
\begin{aligned}
p_n(\eta) &= \frac{\Gamma\left(\frac{n}{2}\right)}{\Gamma\left(\frac{n-1}{2}\right) \sqrt{\pi}} \sin^{n-2}(\arccos \eta) \left| \frac{d \theta}{d \eta} \right| \\
&= \frac{\Gamma\left(\frac{n}{2}\right)}{\Gamma\left(\frac{n-1}{2}\right) \sqrt{\pi}} \left(1 - \eta^2\right)^{(n-3) / 2}
\end{aligned}
\end{equation}
From the form $p_n(\theta) \sim \sin^{n-2} \theta$, we observe that when $n \geq 3$, the maximum probability occurs at $\theta = \frac{\pi}{2}$ (i.e., 90 degrees). Additionally, $\sin^{n-2} \theta$ is symmetric about $\theta = \frac{\pi}{2}$, so its mean is also $\frac{\pi}{2}$. However, this does not fully describe the distribution; we also need to consider the variance.

The variance is given by:
\begin{equation}
\operatorname{Var}_n(\theta) = \frac{\Gamma\left(\frac{n}{2}\right)}{\Gamma\left(\frac{n-1}{2}\right) \sqrt{\pi}} \int_0^\pi \left(\theta - \frac{\pi}{2}\right)^2 \sin^{n-2} \theta \, d\theta
\end{equation}

To obtain an approximate analytical solution, we can use the Laplace method. Expanding \(\ln \sin^{n-2} \theta\) around \(\theta = \frac{\pi}{2}\), we get:
\begin{equation}
\ln \sin^{n-2} \theta = \frac{2-n}{2} \left(\theta - \frac{\pi}{2}\right)^2 + \mathcal{O}\left(\left(\theta - \frac{\pi}{2}\right)^4\right)
\end{equation}

Thus,
\begin{equation}
\sin^{n-2} \theta \approx \exp \left[-\frac{n-2}{2} \left(\theta - \frac{\pi}{2}\right)^2\right]
\end{equation}

From this approximation, we can approximately consider that \(\theta\) follows a normal distribution with mean \(\frac{\pi}{2}\) and variance \(\frac{1}{n-2}\). As \(n\) becomes larger, the variance approximates to \(\frac{1}{n-2}\), indicating that the variance decreases as \(n\) increases.

\subsection{Proof of THEOREM 2}
\label{proof2}
Inspired by~\cite{club}, we calculate the gap between $\mathbb{I}_{\mathrm{upper}}$ and $\mathbb{I}(\boldsymbol{x} ; \boldsymbol{y})$ :
\begin{equation}
\begin{aligned}
\tilde{\Delta} & := \mathbb{I}_{\mathrm{upper}}(\mathbf{E}^g ; \mathbf{E}^q) - \mathrm{I}(\mathbf{E}^g ; \mathbf{E}^q) \\
& = \mathbb{E}_{p(\mathbf{E}^g, \mathbf{E}^q)}\left[\log q_\theta(\mathbf{E}^q \mid \mathbf{E}^g)\right] - \mathbb{E}_{p(\mathbf{E}^g)} \mathbb{E}_{p(\mathbf{E}^q)}\left[\log q_\theta(\mathbf{E}^q \mid \mathbf{E}^g)\right] \\
& \quad - \mathbb{E}_{p(\mathbf{E}^g, \mathbf{E}^q)}[\log p(\mathbf{E}^q \mid \mathbf{E}^g) - \log p(\mathbf{E}^q)] \\
& = \left[\mathbb{E}_{p(\mathbf{E}^q)}[\log p(\mathbf{E}^q)] - \mathbb{E}_{p(\mathbf{E}^g) p(\mathbf{E}^q)}\left[\log q_\theta(\mathbf{E}^q \mid \mathbf{E}^g)\right]\right] \\
& \quad - \left[\mathbb{E}_{p(\mathbf{E}^g, \mathbf{E}^q)}[\log p(\mathbf{E}^q \mid \mathbf{E}^g)] 
- \mathbb{E}_{p(\mathbf{E}^g, \mathbf{E}^q)}\left[\log q_\theta(\mathbf{E}^q \mid \mathbf{E}^g)\right]\right] \\
& = \mathbb{E}_{p(\mathbf{E}^g) p(\mathbf{E}^q)}\left[\log \frac{p(\mathbf{E}^q)}{q_\theta(\mathbf{E}^q \mid \mathbf{E}^g)}\right] - \mathbb{E}_{p(\mathbf{E}^g, \mathbf{E}^q)}\left[\log \frac{p(\mathbf{E}^q \mid \mathbf{E}^g)}{q_\theta(\mathbf{E}^q \mid \mathbf{E}^g)}\right] \\
& = \mathbb{E}_{p(\mathbf{E}^g) p(\mathbf{E}^q)}\left[\log \frac{p(\mathbf{E}^g) p(\mathbf{E}^q)}{q_\theta(\mathbf{E}^q \mid \mathbf{E}^g) p(\mathbf{E}^g)}\right] \\
& - \mathbb{E}_{p(\mathbf{E}^g, \mathbf{E}^q)}\left[\log \frac{p(\mathbf{E}^q \mid \mathbf{E}^g) p(\mathbf{E}^g)}{q_\theta(\mathbf{E}^q \mid \mathbf{E}^g) p(\mathbf{E}^g)}\right] \\
& = \mathrm{KL}\left(p(\mathbf{E}^g) p(\mathbf{E}^q) \| q_\theta(\mathbf{E}^g, \mathbf{E}^q)\right) - \mathrm{KL}\left(p(\mathbf{E}^g, \mathbf{E}^q) \| q_\theta(\mathbf{E}^g, \mathbf{E}^q)\right).
\end{aligned}
\end{equation}

Therefore, $\mathbb{I}_{\mathrm{upper}}(\mathbf{E}^g ; \mathbf{E}^q)$ is an upper bound of $\mathbb{I}(\mathbf{E}^g ; \mathbf{E}^q)$ if and only if $\mathrm{KL}\left(p(\mathbf{E}^g) p(\mathbf{E}^q) \| q_\theta(\mathbf{E}^g, \mathbf{E}^q)\right) \geq \operatorname{KL}\left(p(\mathbf{E}^g, \mathbf{E}^q) \| q_\theta(\mathbf{E}^g, \mathbf{E}^q)\right)$.

\begin{table*}[]
    \caption{Efficiency Analysis. We statistics the memory and the running time per epoch for each model.}
    \setlength\tabcolsep{4pt}
    \label{effiency}
    \begin{tabular}{cccccccccccc}
    \toprule 
    \multirow{2}{*}{ Dataset } & \multirow{2}{*}{ Metric } & \multicolumn{3}{c}{ General CF model } & \multicolumn{7}{c}{ Multimodal model } \\
    \cmidrule(l){3-12}
    & & BPR & LightGCN & VBPR & MMGCN  & LATTICE & FREEDOM & \shortname$_d$ & \shortname$_c$ & \shortname$_b$ & \shortname$_a$ \\
    \midrule \multirow{2}{*}{ Baby } & Memory (GB) & 1.59 & 1.69 & 1.89 & 2.69 & 4.53 & 2.13 & 2.54 &  3.42 & 2.54 &  3.42 \\
    & Time (s/epoch) & 0.86 & 1.81 & 1.04 & 6.35  & 3.42 & 2.58 & 4.58 & 18.81 & 3.75 &  16.04  \\
    \midrule \multirow{2}{*}{ Sports } & Memory (GB) & 2.00 & 2.24 & 2.71 & 3.91 & 19.93 & 3.34 & 3.94 &  8.83 & 3.94 &  8.83  \\
    & Time (s/epoch) & 1.24 & 3.74 & 1.67 & 20.88  & 13.59 & 4.52 & 16.13 & 60.97 &  13.75 &  58.81  \\
    \midrule \multirow{2}{*}{ Clothing } & Memory (GB) & 2.16 & 2.43 & 3.02 & 4.24 & 28.22 & 4.15 &  4.69 &  13.78 & 4.69 &  13.78  \\
    & Time (s/epoch) & 1.52 & 4.58 & 2.02 & 26.13  & 23.96 & 5.74 & 23.59 & 59.25  & 21.41 &  57.35  \\
    \bottomrule
    \end{tabular}
\end{table*}

\subsection{Compared Methods}
To ensure that our comparison models are diverse, we compare not only collaborative filtering models, but also existing multimodal models, which include classical, as well as recently published ones (for presentation purposes, we add the publication avenue as well as the year after the model name). For this purpose, We choose the following methods.
\label{compared}
\begin{itemize}
  \item \textbf{BPR}~\cite{bpr} is bayesian personalised ranking, one of the classical recommendation methods. 
  \item \textbf{LightGCN}~\cite{lightgcn} simplifies the components in GCN, lightweighting the model while gaining better performance. 
  \item \textbf{LayerGCN}~\cite{layergcn} alleviates the oversmoothing problem and sparsifies the user-item graph.
  \item \textbf{VBPR}~\cite{vbpr} incorporates visual information on the basis of BPR, fusing latent representation with visual features extracted by convolutional neural networks for recommendation. 
  \item \textbf{MMGCN}~\cite{MMGCN} process the textual user-item graph and the visual user-item graph separately with GCN and fuse the information from both.
  \item \textbf{DualGNN}~\cite{dualgnn} fuses the user representation obtained from the user-item graph and the user-user co-occurrence graph as the final representation. 
  \item \textbf{LATTICE}~\cite{LATTICE} obtains the item-item graph by KNN~(K-Nearest Neighbors) algorithm and fuses its obtained representation with those of the user-item graph.~\looseness=-2
  \item \textbf{SLMRec}~\cite{SLMRec} introduces self-supervised learning, treating ID, vision, text, and sound as one modality each, and constructing separate modal-unique graphs for each modality. 
  \item \textbf{MICRO}~\cite{MICRO} is an extension of LATTICE that mines the latent structure between items by learning an item-item graph and obtaining the items' multimodal representation by contrastive fusion. 
  \item \textbf{PAMD}~\cite{MICRO} is a pretraining framework that learns modality-common and modality-specific representations for recommendation through a disentangled encoder and contrastive learning.
  \item \textbf{BM3}~\cite{BM3} utilizes self-supervised learning to do intra-modal alignment and inter-modal alignment. \looseness=-1
  \item \textbf{MMSSL}~\cite{MMSSL} propose cross-modal contrastive learning that jointly preserves the commonality of cross-modal semantics and the diversity of user preferences.
  \item \textbf{FREEDOM}~\cite{freedom} is the plus version of LATTICE, specifically, denoising the user-item graph, and freezing the item-item graph.
  \item \textbf{MGCN}~\cite{MGCN} adopts attention
layers to capture the modalities’ importance and design a self-supervised auxiliary task that aims to maximize the mutual information between modal and behavioral features. 
  \item \textbf{LGMRec}~\cite{guo2023lgmrec} uses a hypergraph neural network to extract various modal information deeply.
\end{itemize}

\vspace{3pt}
\noindent

\subsection{Negative $\ell_2$ Distance}
\label{l2}
We have already introduced capable mutual information minimization, but the time complexity is relatively high. Inspired by~\cite{SimMMDG}, we provide a simpler solution, negative $\ell_2$ distance, for use when resources are constrained. To achieve that purpose, we utilize a negative $\ell_2$ loss function as follows:
\begin{equation}
\mathcal{L}_{m\_\text{dis}}=- \left\|\mathbf{E}_q^m-\mathbf{E}_g^m\right\|_2^2,
\end{equation}
where $|m|$ represents the number of modalities, $\mathbf{E}_q^m$ denotes the modal-unique features, and $\mathbf{E}_g^m$ represents the modal-generic features.

\end{document}